\documentclass[twocolumn,superscriptaddress,showpacs,amsmath,amssymb,pra,longbibliography]{revtex4-1}
\usepackage{dcolumn,bm,book tabs,comment,natbib,isomath,mathtools}
\usepackage[pdftex]{graphicx}
\usepackage[usenames,dvipsnames]{color}
\usepackage[colorinlistoftodos]{todonotes}
\usepackage{colortbl}
\usepackage{cancel}

\graphicspath{{../2014-12-14_vu_141213/}}

\definecolor{URLCOL}{rgb}{0,0.52,0.83} % external link
\definecolor{LINKCOL}{rgb}{0.05,0.5,0} % internal link
\definecolor{CITECOL}{rgb}{0.25,0,0.48} % link to bibliography

\usepackage[pdftex,bookmarks,breaklinks,bookmarksopen,bookmarksnumbered,colorlinks,linkcolor=LINKCOL,linktocpage,citecolor=CITECOL,urlcolor=URLCOL,pdfpagemode=UseOutlines,pdftex]{hyperref}

\definecolor{TITLECOL}{rgb}{0.1,0.2,0.7} % title
\definecolor{SECOL}{rgb}{0.1,0.2,0.7} % section
\definecolor{CONTENTSCOL}{rgb}{0.1,0.2,0.7} % table of contents
\definecolor{SSECOL}{rgb}{0.25,0,0.48} % subsection
\definecolor{SSSECOL}{rgb}{0.2,0.08,0.53} % subsubsection
\definecolor{FINCOL}{rgb}{0.01,0.3,0.07}

\def\coloredtitle#1{\title{\textcolor{TITLECOL}{#1}}} % title color
\def\coloredauthor#1{\author{\textcolor{CITECOL}{#1}}} % author color

\definecolor{URLCOL}{rgb}{0,0.17,0.43} % external link
\definecolor{LINKCOL}{rgb}{0.05,0.4,0} % internal link
\definecolor{CITECOL}{rgb}{0.35,0,0.48} % link to bibliography

\def\Eqref#1{Eq.~\eqref{#1}}
\def\Secref#1{Section~\ref{#1}}
\def\Figref#1{Fig.~\ref{#1}}
\def\Ref#1{Ref.~\cite{#1}}

\def\sec#1{\section{\textcolor{SECOL}{#1}}}
\def\ssec#1{\subsection{\textcolor{SSECOL}{#1}}}
\def\sssec#1{\subsubsection{\textcolor{SSSECOL}{#1}}}

\definecolor{lightgray}{gray}{0.8}

\def\figwidth{7cm}
\def\bea{\begin{eqnarray}}
\def\eea{\end{eqnarray}}
\def\ben{\begin{equation}}
\def\een{\end{equation}}
\def\benu{\begin{enumerate}}
\def\enu{\end{enumerate}}

\def\bei{\begin{itemize}}
\def\eei{\end{itemize}}
\def\beit{\begin{itemize}}
\def\eit{\end{itemize}}
\def\benu{\begin{enumerate}}
\def\enu{\end{enumerate}}

\def\n{n}

\def\sss{\scriptscriptstyle\rm}

\def\g{_\gamma}

\def\l{^\lambda}

\def\1var{(\bx_1...\bx\N)}

\def\half{\frac{1}{2}}

\def\bx{{x}}

\def\s{_{\sss S}}

\def\N{_{\sss N}}

\def\I{^{\rm I}}

\def\sph_int{ {\int d^3 r}}

\def\W{^{\rm W}}

\def\ML{^{\rm ML}}

\def\K{\matrixsym{K}}
\def\I{\matrixsym{I}}
\def\f{\vectorsym{f}}
\def\i{{\rm i}}
\DeclareMathOperator{\erf}{erf}

\makeatletter
\newcommand{\Spvek}[2][r]{%
  \gdef\@VORNE{1}
  \left(\hskip-\arraycolsep%
    \begin{array}{#1}\vekSp@lten{#2}\end{array}%
  \hskip-\arraycolsep\right)}

\def\vekSp@lten#1{\xvekSp@lten#1;vekL@stLine;}
\def\vekL@stLine{vekL@stLine}
\def\xvekSp@lten#1;{\def\temp{#1}%
  \ifx\temp\vekL@stLine
  \else
    \ifnum\@VORNE=1\gdef\@VORNE{0}
    \else\@arraycr\fi%
    #1%
    \expandafter\xvekSp@lten
  \fi}
\makeatother

\begin{document}

\coloredtitle{
Understanding Kernel Ridge Regression: \\
Common behaviors from simple functions to density functionals
}

\coloredauthor{Kevin Vu}
\affiliation{Department of Physics and Astronomy, University of California, Irvine, CA 92697}

\coloredauthor{John Snyder}
\affiliation{Machine Learning Group, Technical University of Berlin, 10587 Berlin, Germany}
\affiliation{Max Planck Institute of Microstructure Physics, Weinberg 2, 06120 Halle (Saale), Germany}

\coloredauthor{Li Li}
\affiliation{Department of Physics and Astronomy, University of California, Irvine, CA 92697}

\coloredauthor{Matthias Rupp}
\affiliation{Department of Chemistry, University of Basel, Klingelbergstr. 80, 4056 Basel, Switzerland}

\coloredauthor{Brandon F. Chen}
\affiliation{Department of Chemistry, University of California, Irvine, CA 92697}
\coloredauthor{Tarek Khelif}
\affiliation{Department of Chemistry, University of California, Irvine, CA 92697}

\coloredauthor{Klaus-Robert M{\"u}ller}
\affiliation{Machine Learning Group, Technical University of Berlin, 10587 Berlin, Germany}
\affiliation{Department of Brain and Cognitive Engineering, Korea University,
Anam-dong, Seongbuk-gu, Seoul 136-713, Korea}

\coloredauthor{Kieron Burke}
\affiliation{Department of Physics and Astronomy, University of California, Irvine, CA 92697}
\affiliation{Department of Chemistry, University of California, Irvine, CA 92697}

\date{\today}

\begin{abstract}
Accurate approximations to density functionals
have recently
been obtained via machine learning (ML). 
By applying ML to a simple function of one variable without any random sampling,
we extract the qualitative dependence of errors on hyperparameters.
We find universal features of the behavior in extreme limits, including
both very small and very large length scales, and the noise-free limit.
We show how such features arise in ML models of density functionals.
\end{abstract}

%\pacs{}

\maketitle

%--------------------------------------------%

\sec{Introduction}
\label{Introduction}

Machine learning (ML) is a
powerful data-driven method for learning patterns in high-dimensional spaces
via induction, and has had widespread success in many fields
including more recent applications in
quantum chemistry and materials science
\cite{MMRT01, K01,S02, I07, BPKC10, RTML12, PHSR12, HMBF13, SGBS14}.
Here we are interested in applications of
ML to  construction of density 
functionals \cite{SRHM12,SRHB13,LSPH14,SMBM13,SRMB15}, which have focused so far on
approximating the kinetic energy (KE) of non-interacting electrons.
An accurate, general approximation to this could make orbital-free DFT
a practical reality.

However, ML methods have been developed within the areas of statistics and
computer science, and have been applied to a huge variety of data, including 
neuroscience, image and text processing, and robotics \cite{B06}.
Thus, they are quite general
and have not been tailored to account for specific details of the quantum
problem.  For example, it was found that a standard method, kernel ridge
regression, could yield excellent results for the KE functional,
while never yielding accurate functional derivatives.  The development of
methods for bypassing this difficulty has been important for ML in general \cite{SMBM13}.

ML provides a whole suite of tools for analyzing data, fitting 
highly non-linear functions, and dimensionality reduction \cite{HTF09}.
More importantly in this context, ML provides
a completely different way of thinking about
electronic structure. 
The traditional {\em ab-initio} approach \cite{L09} to electronic structure involves 
deriving carefully constructed approximations to solving the Schr\"odinger
equation, based on physical intuition, exact conditions and asymptotic
behaviors \cite{DG90}.
On the other hand, ML learns by example. Given a set of training data,
ML algorithms learn via induction to predict 
new data. ML provides limited interpolation over a specific class of 
systems for which training data is available.

A system of $N$ interacting electrons with some
external potential is characterized by a $3N$ coordinate wavefunction,
which becomes computationally demanding for large $N$. 
In the mid 1960's, Hohenberg and Kohn proved a one-to-one correspondence 
between the external potential of a quantum system and its one-electron 
ground-state density \cite{HK64}, showing that all properties are functionals of
the ground-state density alone, which can in principle be found from
a single Euler equation for the density.
Although these 
fundamental theorems of DFT proved the existence of a universal 
functional, essentially all modern calculations use the KS scheme \cite{KS65},
which is much more accurate, because the non-interacting KE
is found exactly by using an orbital-scheme \cite{PGB14}.  This is far faster than
traditional approaches for large $N$, but remains a bottleneck.
If a sufficiently accurate density functional for the non-interacting
electrons could be found, it could increase the size of computationally
tractable systems by orders of magnitude.

The Hohenberg-Kohn theorem guarantees that all properties of
the system can be determined from the electronic density alone. 
The basic tenet of ML is that a pattern must
exist in the data in order for learning to be possible. Thus,
DFT seems an ideal case to apply ML.  ML learns the underlying pattern 
in {\em solutions} to the Schr\"odinger equation, bypassing the need to 
directly solve it.  The HK theorem is a statement concerning the minimal
information needed to do this for an arbitrary one-body potential.

Some of us recently used ML to learn the non-interacting
KE of fermions in a one-dimensional box subject to
smooth external potentials \cite{SRHM12} and of
a one-dimensional model of diatomics where we demonstrated
the ability of ML to break multiple bonds self-consistently via
an orbital-free density functional theory (DFT) \cite{SRHB13}.
Such KE data is effectively {\em noise-free}, since it is generated via 
deterministic reference calculations, by solving the Schr\"odinger equation 
or KS equations numerically exactly. 
(The limited precision of the calculation might be considered
``noise,'' as different implementations might
yield answers differing on the order of machine precision, but this is negligble.)
There is no noise, in the
traditional sense, as is typically associated with experimental data. 
Note that what is considered ``noise'' depends on what
is considered ground truth, i.e., the data to be learned.
In particular, if a single reference method is used, its
error with respect to a universal functional is 
\emph{not} considered noise for the ML model.  A perfect ML model should, at best,
precisely reproduce the single-reference calculation.

\begin{figure}[tb]
\includegraphics[width=\figwidth]{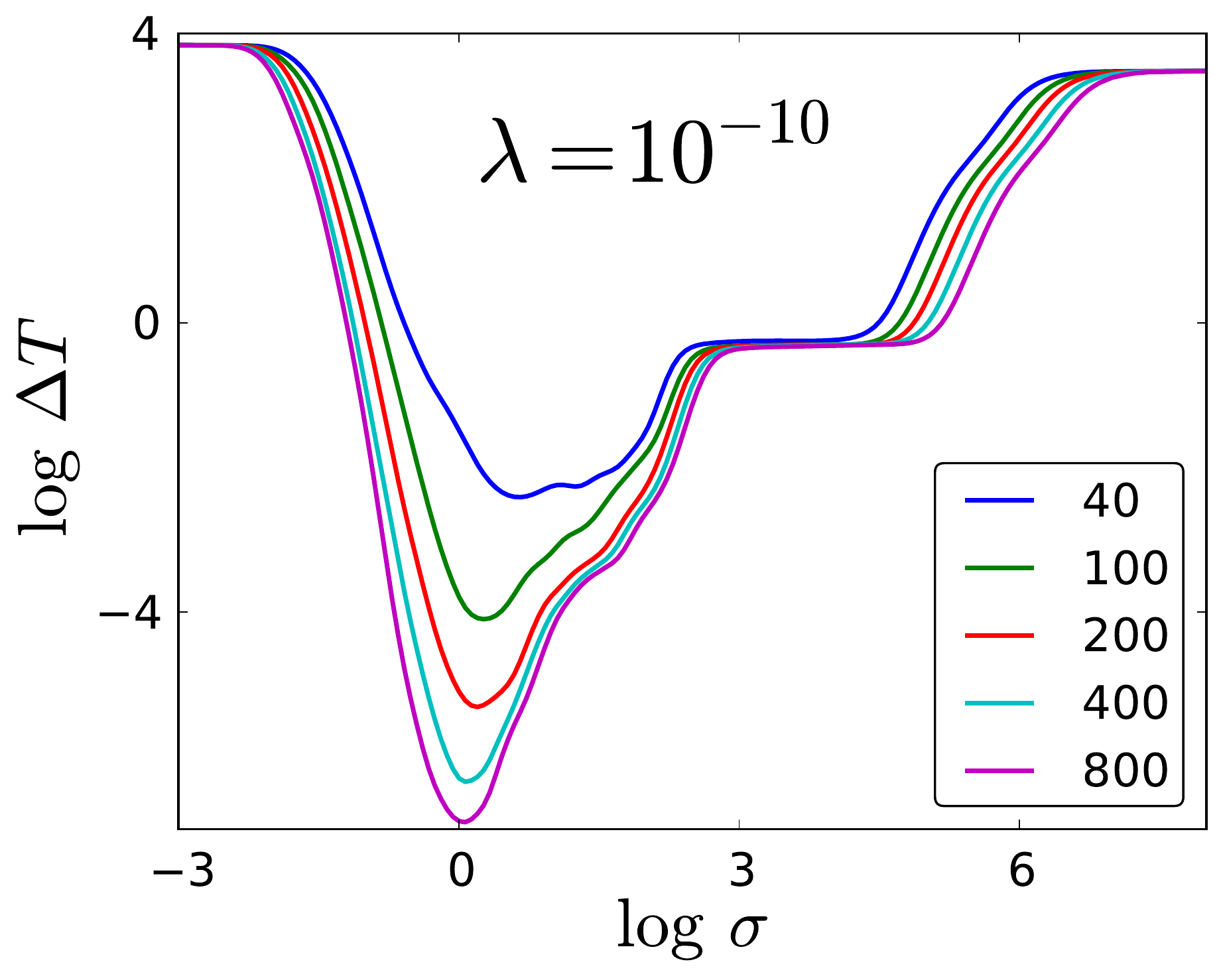}
\caption{The error of the model, $\Delta T$ (Hartree), for the KE of particles in a box (\Secref{dft-section}) as a function of $\sigma$,
for fixed $\lambda=10^{-10}$. $N_T$ values for each curve are given in the legend.}
\label{9-box-delT-NT}
\end{figure}

As an example, in \Figref{9-box-delT-NT} we plot a measure of
the error of ML for the KE of up to 4
noninteracting spinless fermions in a box under 
a potential with 9 parameters (given in detail in~\Ref{SRHM12}), fitted for
different numbers of evenly spaced training densities as a function
of the hyperparameter $\sigma$ (called the length scale), for fixed
$\lambda$ (a hyperparameter called the regularization strength) and several different number of
training points $N_T$. The scale is logarithmic \footnote{
We will use $\log$ to denote $\log_{10}$ here and throughout this work},
so there are large variations in the fitted error. We will give a more in-depth analysis of the model performance on this data set in a later section after we have formally defined the functions and hyperparameters involved, but for now it is still useful to observe the qualitative behaviors that emerge in the figure. Note that the curves assume roughly the same shape for each $N_T$ over the range of $\sigma$, and that they all possess distinct features in different regimes of $\sigma$. 

\begin{figure}[tb]
\includegraphics[width=\figwidth]{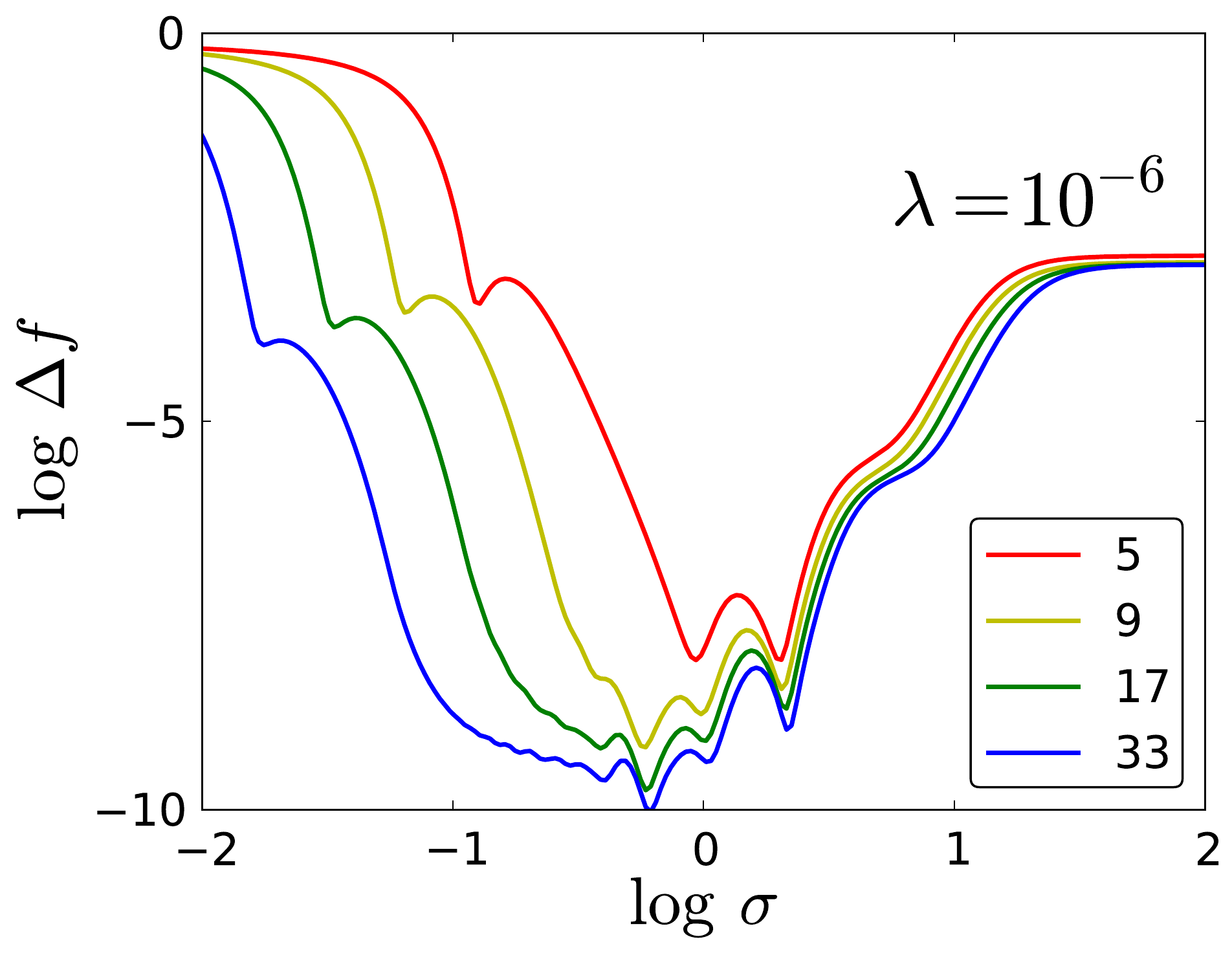}
\caption{Dependence of the model error (as a function
of $\sigma$) when fitting $\cos x$ between $0$ and $1$ (\Secref{Analysis}) for various $N_T$ (shown in the legend) for $\lambda=10^{-6}$.}
\label{delf-Nt}
\end{figure}

To better understand the behavior with respect to hyperparameters seen in Fig 1,
we have chosen in this
paper to apply them to the prototypical regression problem, that of 
fitting a simple function of one coordinate. We also remove all
stochastic elements of the procedure, by considering data points
on uniform grids, defining errors in the continuum limit, etc.
This is shown in Fig. \ref{delf-Nt}, where we plot a measure of the error of ML for a simple function $\cos x$, fitted in the region between 0 and 1, inclusive, for several $N_T$ (represented as values on a grid) as a function
of $\sigma$. Note the remarkable similarity between the features and characteristics of the curves of this figure and those of \Figref{9-box-delT-NT} (like \Figref{9-box-delT-NT} before it, we will give a more in-depth analysis of \Figref{delf-Nt} later).   
We explore the behavior of the fitting error as a function
of the number of training parameters and the hyperparameters that
are used in kernel ridge regression with Gaussian kernel.  We find the landscape to 
be surprisingly rich, and we also find elegant simplicities in 
various limiting cases. After this, we will be able to characterize the behavior of ML for systems like the one shown in \Figref{9-box-delT-NT}.

Looking at \Figref{delf-Nt}, we see that the best results (lowest error) are always obtained from the middle of the
curves, which can become quite flat with enough training data.  Thus, any method
for determining hyperparameters should usually yield a length scale somewhere
in this valley.
For very small length scales, all curves converge to the
same poor result, regardless of the number of training points. On the other hand, notice also the plateau structure
that develops for very large length scales, again with all curves converging to a certain limit.
We show for which ranges of hyperparameters these plateaus emerge and how they can be
estimated.  We also study and explain many of the features of these curves.
To show the value of this study, we then apply the same reasoning to 
the problem that was tackled in \Ref{SRHM12,LSPH14}, which we showcased in \Figref{9-box-delT-NT}.
From the machine learning perspective our study may appear unusual as it considers properties in data and problems that are uncommon. Namely there are  
only a few noise free data points and all are low dimensional. Nevertheless, from the physics point of view the toy data considered reflects very well the essential properties of a highly relevant problem in quantum physics: the machine learning of DFT.

%--------------------------------------------%
\sec{Background}
\label{Background}

In this work, we will first use ML to fit a
very simple function of one variable,
\ben
f(x) = \cos x,
\label{eq:model}
\een
on the interval $x\in[0,1]$. We will focus on exploring the properties of ML for this simple function before proceeding to our DFT cases. We choose a set of $x$-values and corresponding $f(x)$ values
as the ``training data'' for ML to learn from.
In ML, the $x$-values $\{x_j\}$ for $j=1,\dots,N_T$ are
known as {\em features}, and corresponding $f$-values,
$\{f_j\}$, are known as {\em labels}. Here $N_T$ is the
number of training samples.
Usually, ML is applied with considerable random elements, such as in the choice of data points and selection of test data. In our case, we choose evenly spaced training points on
the interval $[0,1]$: $x_j = (j - 1)/(N_T - 1)$ for $j=1,\dots,N_T$.

Using this dataset, we apply kernel ridge regression (KRR), 
which is a non-linear
form of regression with regularization to prevent overfitting
\cite{HTF09}, to fit $f(x)$. The general form of KRR is
\ben
f\ML(x) = \sum_{j=1}^{N_T} \alpha_j k(x, x_j),
\label{modeldef}
\een
where $\alpha_j$ are the {\em weights}, $k$ is the kernel (which is a measure of similarity
between features), and the hyperparameter $\lambda$ controls the strength of the regularization and is linked to the noise level of the learning problem.
We use the Gaussian kernel
\ben
k(x, x') = \exp \left (-(x - x')^2/2\sigma^2 \right ),
\label{kerneldef}
\een
a standard choice in ML that works well for a variety of problems.
The hyperparameter $\sigma$ is the length scale of the Gaussian,
which controls the degree of correlation between training points.

The weights $\alpha_j$ are obtained through the minimization of the cost function
\ben
\mathcal{C}(\boldsymbol{\alpha}) = \sum_{j=1}^{N_T} \left(f\ML(x_j) - f_j\right)^2 + \lambda \, \boldsymbol{\alpha}^T \K \boldsymbol{\alpha},
\label{costfunction}
\een
where 
\ben
\boldsymbol{\alpha} = \left(\alpha_1, \ldots, \alpha_{N_T} \right)^T.
\een
The exact solution is given by
\ben
\boldsymbol{\alpha} = (\K + \lambda\I)^{-1}\f,
\label{weights}
\een
where $\I$ is the $N_T \times N_T$ identity matrix, $\K$ is the kernel matrix with elements $K_{ij}$ = $k(x_i, x_j)$, and
$\f = (f_1, \dots, f_{N_T})^T$.

The two parameters $\lambda$ and $\sigma$ not determined by \Eqref{weights} are called hyperparameters and must be determined from the data (see section~\ref{crossvalidation}).
$\sigma$ can be viewed as the characteristic length scale of the problem being learned (the scale on which changes of $f$ take place), as discernible from the data (and thus dependent on, e.g., the number of training samples).
$\lambda$ controls the leeway the model has to fit the training points. 
For small~$\lambda$, the model has to fit the training points exactly, 
whereas for larger~$\lambda$ some deviation is allowed.
Larger values of $\lambda$ therefore cause the model to be smoother and vary less, i.e., less prone to overfitting. 
This can be directly seen in Gaussian process regression \cite{rw2006}, a related Bayesian ML model with predictions identical to those of KRR.
There, $\lambda$ formally is the variance of the assumed additive Gaussian noise in values of $f$.

KRR is a method of interpolation. Here, we
are mainly concerned with the error of the machine learning approximation (MLA) to $f(x)$ in the {\em interpolation region}, which in
this case is the interval $x\in[0,1]$. As a measure of this error, we define
\ben
\Delta f = \int_0^1 dx \, (f(x) - f\ML(x))^2.
\label{delf}
\een
In the case of the Gaussian kernel, we can expand this and derive the integrals that appear analytically.
To simplify the analytical process, we define
\ben
\Delta f_0 = \int_0^1 dx \, f^2(x),
\label{delf0}
\een
as the benchmark error when $f\ML(x)\equiv 0$. 
For the cosine function in \Eqref{eq:model}, 
\ben
\Delta f_0=\int_0^1 dx \,\cos^2(x)=\frac{1}{2}+\frac{\sin(2)}{4} \approx 0.7273.
\label{delf0cos}
\een

Our goal is to characterize the dependence of the 
performance of the model on the size of the training data set $(N_T)$ and the hyperparameters of the model $(\sigma,\,\lambda)$. For this simple model,
we discuss different regions of qualitative behavior
and derive the dependence of $\Delta f$ for 
various limiting values of these hyperparameters;
we do all of this in the next few sections. In \Secref{dft-section}, we
discuss how these results can be qualitatively
generalized for the problem of using ML to learn the
KE functional for non-interacting fermions in the box
for a limited class of potentials.

%--------------------------------------------%

\sec{Analysis}
\label{Analysis}

We begin by analyzing the structure of $\Delta f$ as a function of $\sigma$ for fixed
$\lambda$ and $N_T$. 
\Figref{delf-Nt} shows $\Delta f$ as a function of $\sigma$ for various $N_T$ while fixing
$\lambda=10^{-6}$. The curves have roughly the same ``valley'' shape for all $N_T$.
The bottom of the valley is an order of magnitude deeper than the walls and may have multiple local minima. These valleys are nearly identical in shape for sufficiently large $N_T$, which indicates that this particular feature arises in a systematic manner as $N_T$ increases.
Moreover, this central valley opens up to the left (i.e., smaller $\sigma$) as $N_T$ increases---
as the training samples become more densely packed, narrower Gaussians are better able to interpolate
the function smoothly.
Thus, with more training samples, a wider range of 
$\sigma$ values will yield an accurate model. 

In addition, a ``cusp'' will begin to appear in the region to the left of the valley, and its general shape remains the same for increasing $N_T$. This is another recurring feature that appears to develop systematically like the valley. For a fixed $N_T$, and starting from the far left, the $\Delta f$ curve begins to decrease monotonically to the right, i.e., as $\sigma$ increases. The cusps mark the first break in this monotonic behavior, as $\Delta f$ increases briefly after reaching this local minimum before resuming its monotonic decrease for increasing $\sigma$ (until this monotonicity is interrupted again in the valley region). The cusps shift to the left as $N_T$ increases, following the trend of the valleys. This indicates that they are a fundamental feature of the $\Delta f$ curves and that their appearances coincide with a particular behavior of the model as it approaches certain $\sigma$ values.
Note that $\Delta f$ decreases nearly monotonically as $N_T$ increases for all $\sigma$. 
This is as expected, since each additional training sample adds another weighted Gaussian, which should
improve the fit. 

\begin{figure}[tb]
\includegraphics[width=\figwidth]{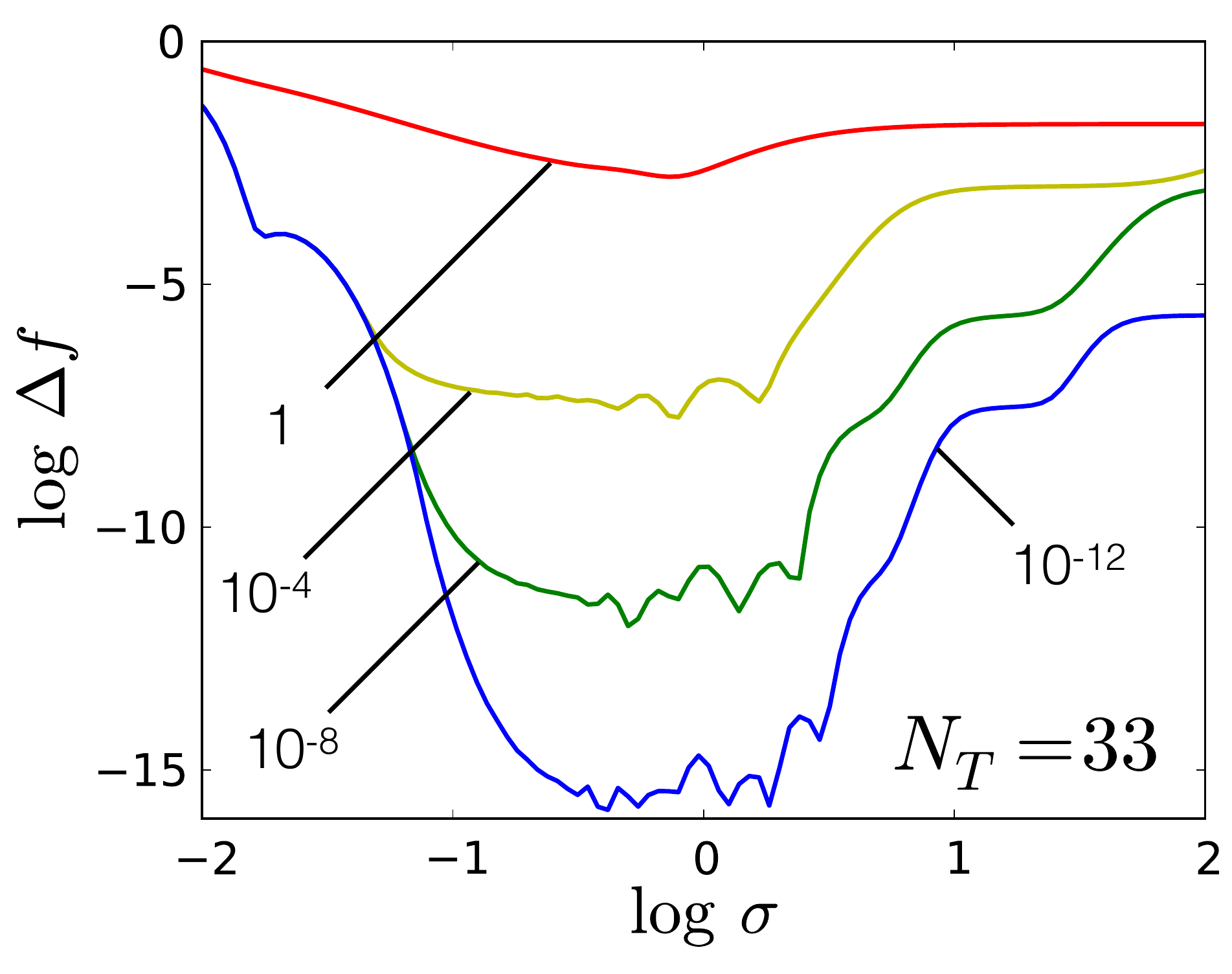}
\caption{Dependence of the model error (as a function
of $\sigma$) for various $\lambda$ with $N_T=33$. The labels give the value of $\lambda$ for each curve.}
\label{delf-lambda-g1}
\end{figure}

Fig. \ref{delf-lambda-g1} again shows $\Delta f$ as a function of $\sigma$, but for various $\lambda$
with $N_T$ fixed at 33.
As $\lambda$ decreases, $\Delta f$ again
decreases nearly monotonically and the central region opens up to the right (i.e., larger $\sigma$). Note that the curves for each $\lambda$ coincide up to a certain $\sigma$ before they split off from the rest, with the lower $\lambda$-valued curves breaking off further along to the right than those with larger $\lambda$.
This shows a well-known phenomenon, namely that regularization strength $\lambda$ and kernel length scale $\sigma$ both give rise to regularization and smoothing \cite{SSM98b}.
Additionally, we observe the emergence of ``plateau''-like structures on the right. These will
be explored in detail in \Secref{regionIII}.

%--------------------------------------------%

\ssec{Regions of qualitative behavior}

\begin{figure}[tb]
\includegraphics[width=\figwidth]{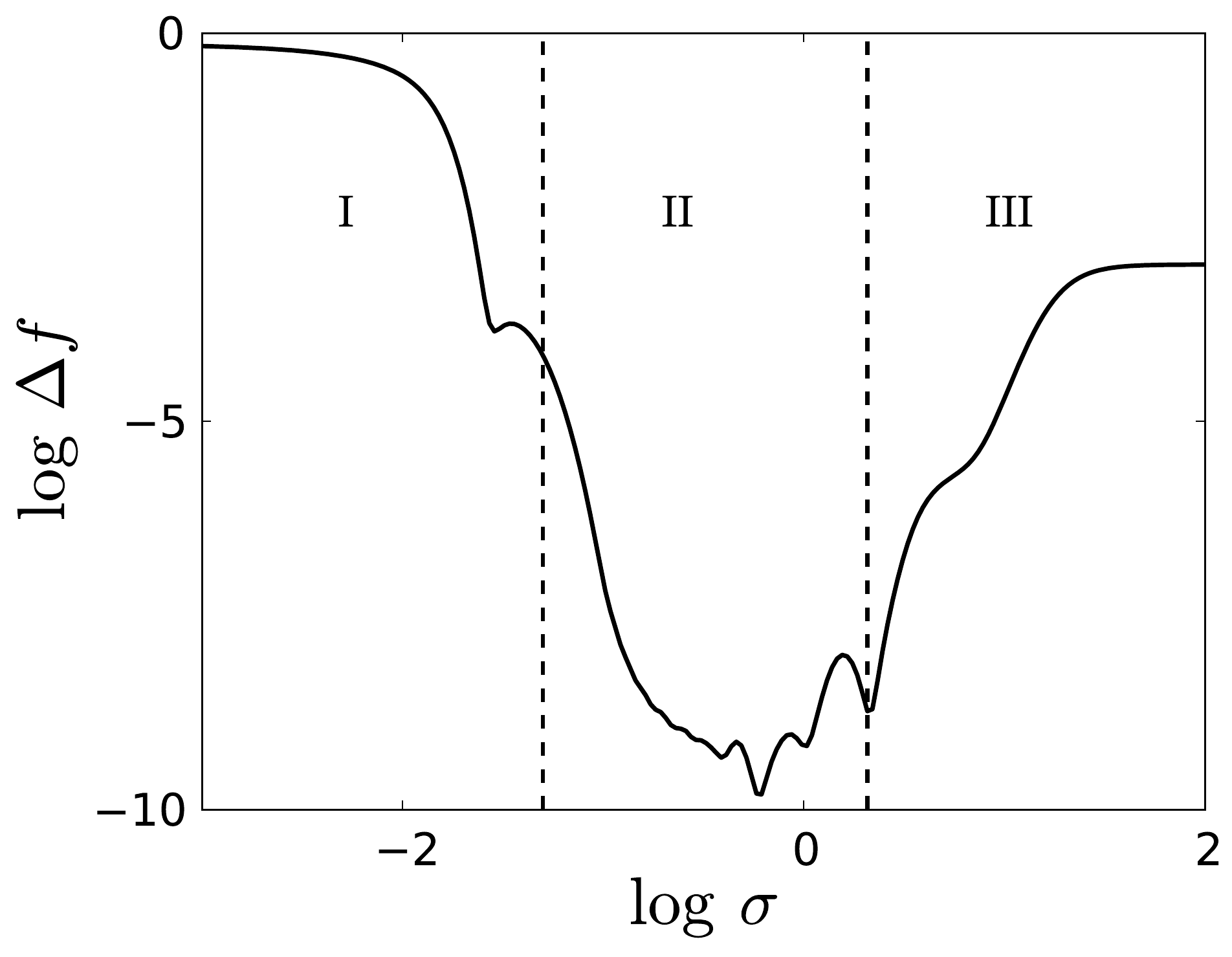}
\caption{Model error $\Delta f$ as a function of $\sigma$ for $N_T=20$ and $\lambda=10^{-6}$. We divide the range of $\sigma$
into three qualitatively distinct regions I, II and III.
The boundaries between the regions are given by the
vertical dashed lines.}
\label{regions}
\end{figure}

In \Figref{regions}, we plot $\Delta f$ as a 
function of $\sigma$ for fixed $\lambda$ 
and $N_T$. 
The three regions labeled I, II, and III denote areas of 
distinct qualitative behavior. They are 
delineated by two arbitrary boundaries we denote by 
$\sigma_s$ ($s$ for small, between I and II) and $\sigma_l$ ($l$ for large, between
II and III). 
In region I,
$\Delta f$ decreases significantly as $\sigma$ increases. The region ends
when there is significant
overlap between neighboring Gaussians (i.e., when
$k(x_j,x_{j+1})$ is no longer small).
Region II is a ``valley'' where the global minimum 
for $\Delta f$ resides.
Region III begins where the valley ends and is populated 
by ``plateaus'' that correspond to $f\ML(x)$ assuming a polynomial form (see \Secref{regionIII}).
In the following sections, we examine each region
separately. In particular, we are interested in the 
asymptotic behavior of $\Delta f$ with respect to $N_T$, $\sigma$ and $\lambda$.

%--------------------------------------------%

\sssec{Length scale too small}

The ML model for $f(x)$, given in \Eqref{modeldef},
is a sum of weighted Gaussians
centered at the training points, where the weights $\alpha_j$
are chosen to best reproduce the unknown $f(x)$.
\begin{figure}[tb]
\includegraphics[width=\figwidth]{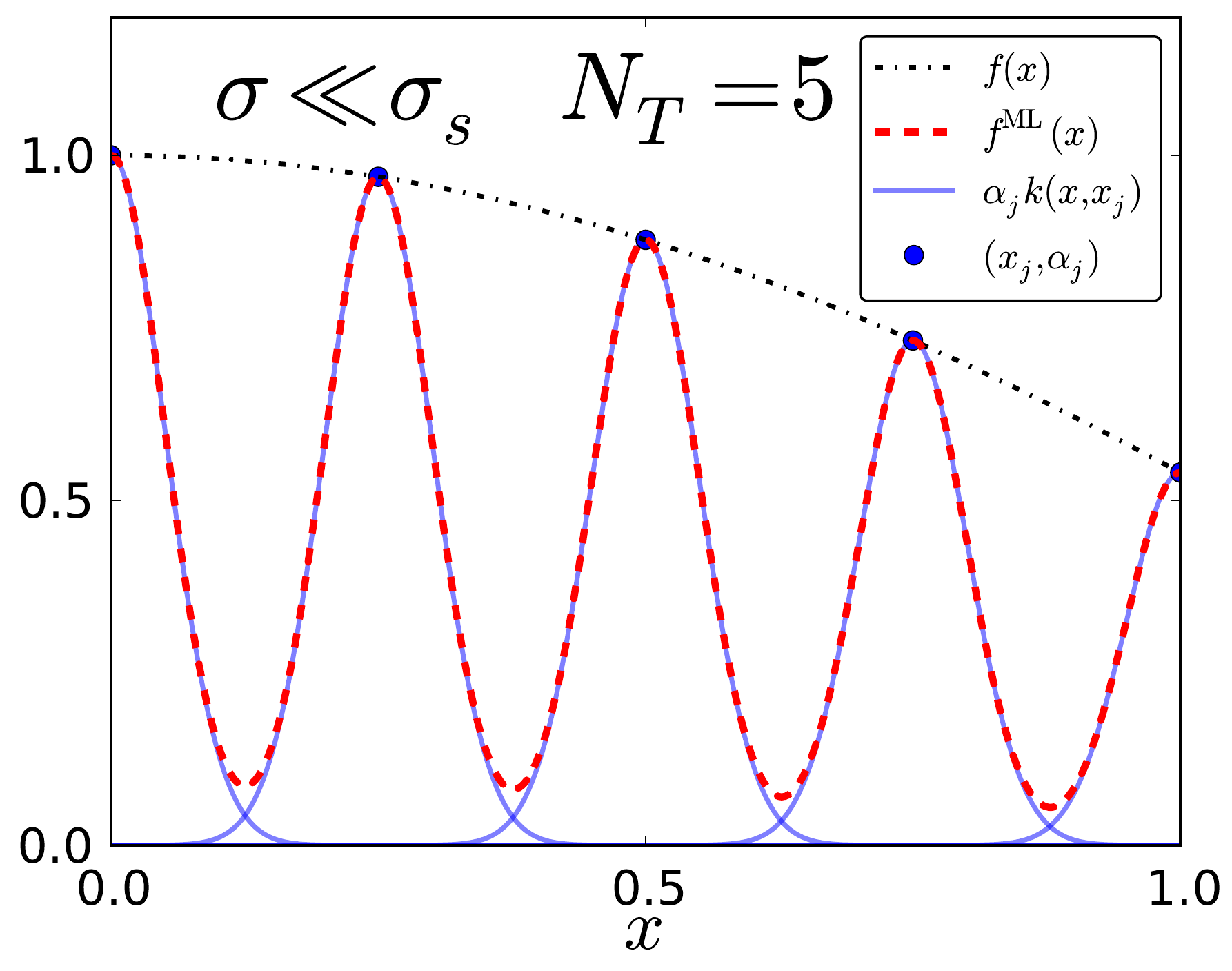}
\caption{Comparison of the function $f(x)$ (black dot-dashed) and the ML model $f\ML(x)$ (red dashed), for 
$N_T=5$, $\sigma=0.05 \ll \sigma_s$ ($\sigma_s=0.2$), and $\lambda=10^{-6}$.
When summed, the weighted Gaussians, $\alpha_j k(x, x_j)$ (blue solid), give $f\ML(x)$. The blue dots show the location of the training points and the value of the corresponding weights.
In this case, the model is in the ``comb'' region, when
$\sigma \ll \sigma_s$. The width of the Gaussians is much smaller than the distance $\Delta x$
between adjacent training points, and so the model cannot reproduce the exact function.}
\label{comb-region}
\end{figure}
\Figref{comb-region} shows what happens when the width
of the Gaussian kernel is too small---the model is incapable
of learning $f(x)$. We call this the ``comb'' region,
as the shape of $f\ML(x)$ arising from the narrow Gaussians resembles a comb.
In order for $f\ML(x)$ to accurately fit $f(x)$,
the weighted Gaussians must have significant overlap.
This begins when $\sigma$ is on the order of the distance between adjacent training points. A corresponding general heuristic is to use a multiple (e.g., four times) of the {\em median nearest neighbor distance} over the training set \cite{SRHB13}.  For equally spaced training data in one dimension, this is $\Delta x \approx 1/N_T$, so we define 
\ben
\sigma_s = 1/N_T
\label{lsL}
\een
to be the boundary between regions I and II.
\begin{figure}[tb]
\includegraphics[width=\figwidth]{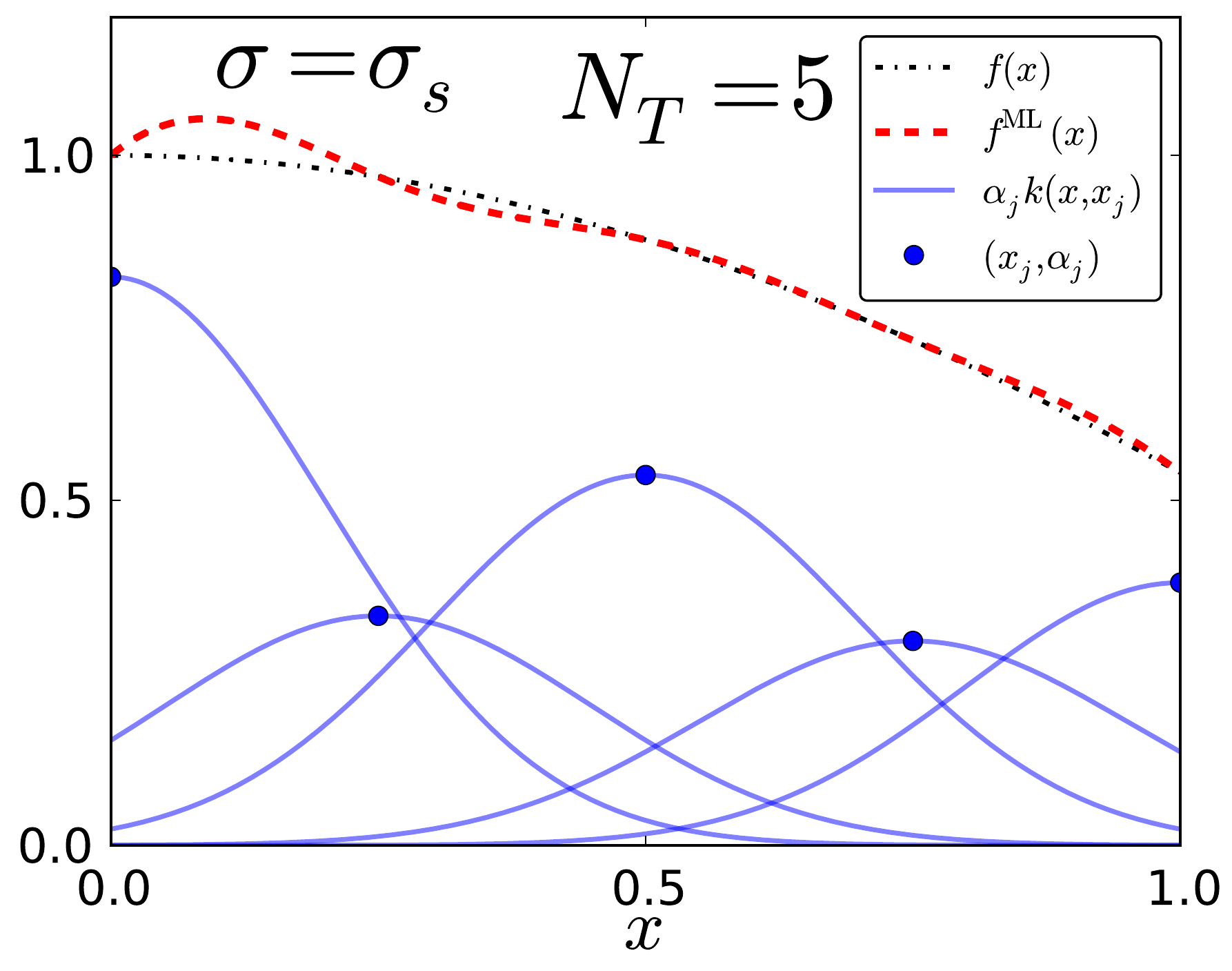}
\caption{Same as \Figref{comb-region}, but for 
$\sigma=\sigma_s=0.2$. Here,
the model is in region II, the optimum region for the model.
The error in the model is very small for all $x$ in
the interpolation region. The width of the Gaussians
is comparable to the size of the interpolation region,
and the weights are large.}
\label{left-bdry}
\end{figure}
In \Figref{left-bdry}, as the overlap between neighboring Gaussians becomes
significant the model is able to reproduce the model
well but still with significant error.
Note that the common heuristics of choosing the length scale in radial basis function networks \cite{MD89} are very much in line with this finding.
In the comb region,
$\Delta f$ decreases as $\sigma$ increases 
in a characteristic way as the Gaussians begin to
fill up the space between the training points.
For $\lambda\to 0$, the weights are approximately given  
as the values of the function at the corresponding 
training points:
\ben
\alpha_j \approx f_j, \quad \sigma \ll \sigma_s.
\een
Thus, for small $\sigma$, the weights are independent
of $\sigma$. Let $\Delta f_{\sigma \ll \sigma_s}$ be
the error of the model when $\alpha_j = f(x_j)$.
\begin{figure}[tb]
\includegraphics[width=\figwidth]{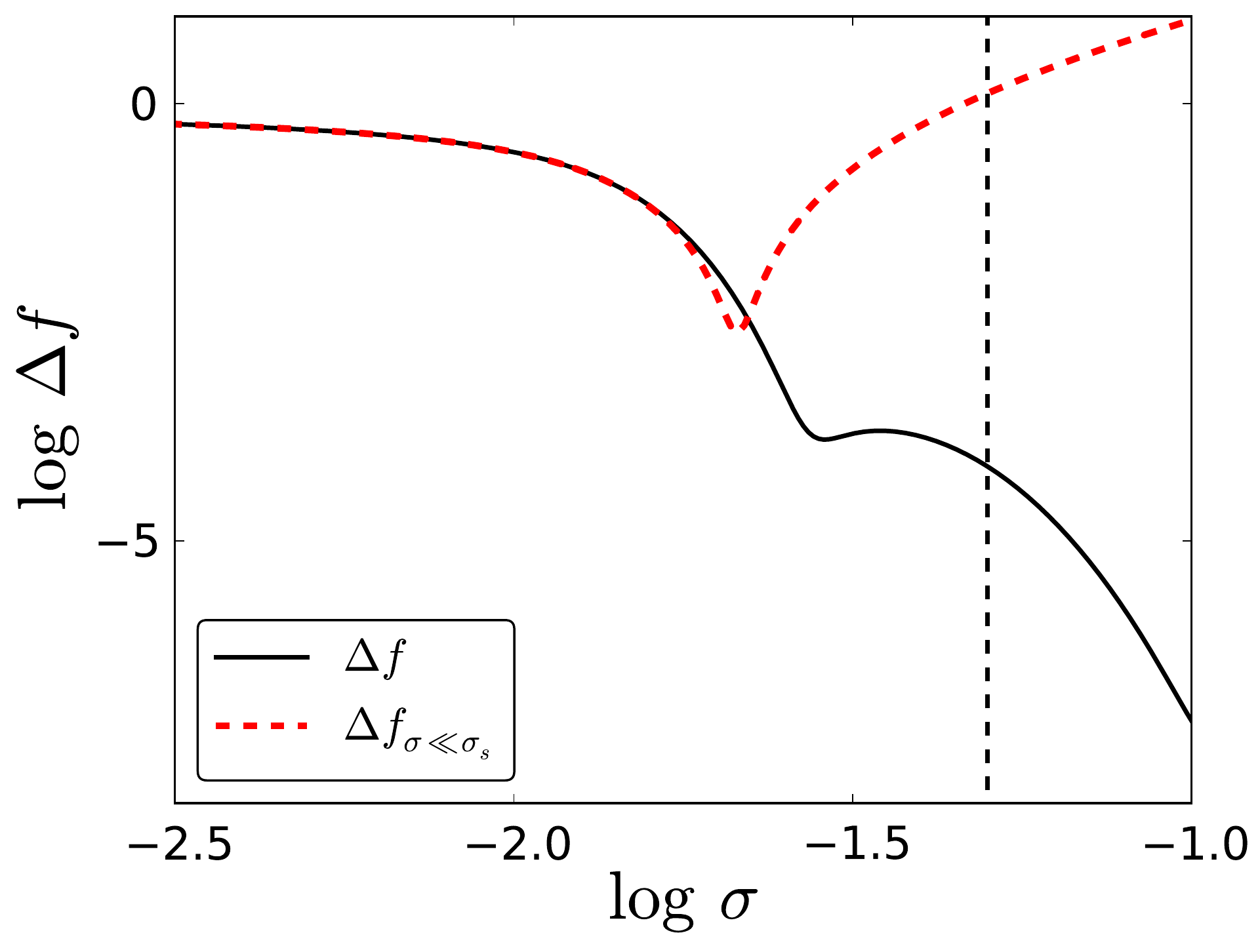}
\caption{Same as \Figref{regions}, but comparing $\Delta f$ against our expansion of $\Delta f$ for $\sigma \ll \sigma_s$ (blue dashed), given in \Eqref{delf-sigma-ll-sigmaL}. The vertical dashed line shows the boundary $\sigma_s$ between regions I and II. The expansion breaks down before we reach $\sigma_s$, as the approximation that $\alpha_j \approx f(x_j)$ is no longer valid.
}
\label{delf-small-sigma}
\end{figure}
This approximation, shown in \Figref{delf-small-sigma},
captures the initial decrease of $\Delta f$ as $\sigma$
increases, but breaks down before we reach $\sigma_s$.
The qualitative nature of this decay is 
independent of the type of function $f(x)$, but its 
location and scale will depend on the specifics.

As $\sigma\to 0$ (for fixed $\lambda$ and $N_T$), 
$f\ML(x)$ becomes the sum of 
infinitesimally narrow Gaussians.
Thus, in this limit, the error in the model becomes
\bea
\lim_{\sigma\to 0}\Delta f &=& \Delta f_0.
\label{limdelf-sigma-0}
\eea
Note that this limit is independent of $\lambda$ and $N_T$.

\begin{figure}[tb]
\includegraphics[width=\figwidth]{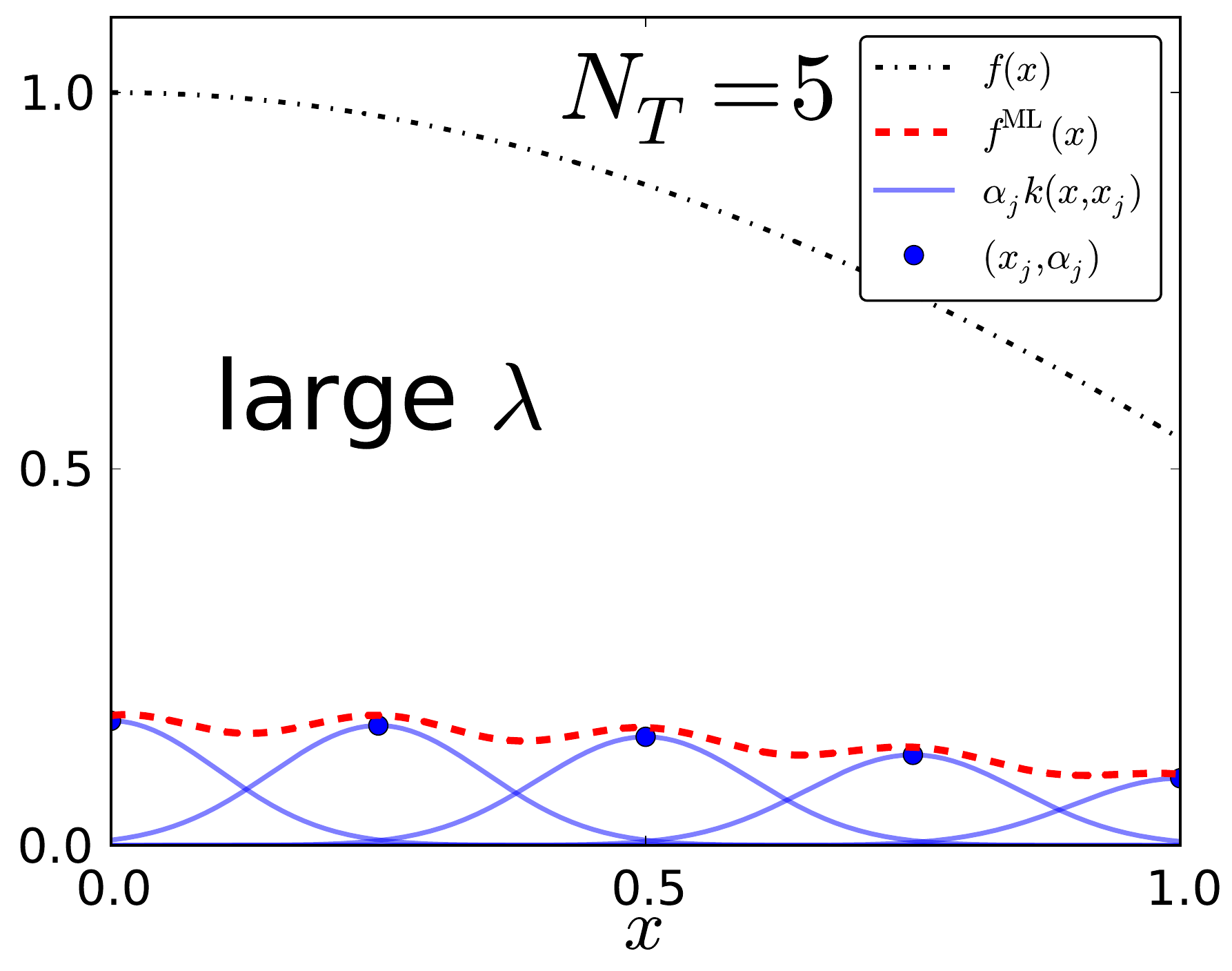}
\caption{Same as \Figref{comb-region}, but for $\sigma=0.1$, and $\lambda=5$.
In this case, known as over-regularization, $\lambda$ is too large, forcing the magnitude of the weights $\alpha_j$ to be small and preventing the model from fitting $f(x)$.
}
\label{over-regularized}
\end{figure}
\Figref{over-regularized} shows what happens when the
regularization becomes too strong. (Although shown for $\sigma$
in region I, the qualitative behavior is the same for any $\sigma$.) The regularization term
in \Eqref{costfunction} forces the magnitude of the weights
to be small, preventing $f\ML(x)$ from fitting 
$f(x)$. As $\lambda\to\infty$,
the weights are driven to zero, and 
so we obtain the same limit as in \Eqref{limdelf-sigma-0}:
\ben
\lim_{\lambda\to\infty} \Delta f = \Delta f_0.
\label{limdelf-lambda-infty}
\een

%--------------------------------------------%

\sssec{Length scale too large}
\label{regionIII}

We define the boundary $\sigma_l$ between regions II and III as the last local minimum of $\Delta f$ (with respect to $\sigma$). Thus, in region III (see Figs.~\ref{delf-Nt} and \ref{delf-lambda-g1})
$\Delta f$ is monotonically increasing.
As $\sigma$ becomes large, the kernel functions become wide and flat over the interpolation region,
and in the limit $\sigma\to\infty$, $f\ML(x)$ is approximately a constant over $x\in[0,1]$. 
For small $\lambda$, the optimal constant will be the average value over the training data
\ben
\lim_{\sigma,1/\lambda\to\infty} \hat f(x) = \frac{1}{N_T} \sum_{j=1}^{N_T} f(x_j).
\een
Note that the order of limits is important here: first $\sigma\to\infty$, then $\lambda\to0$.
If the order is reversed, $f\ML(x)$ becomes the best polynomial fit of order $N_T$. We will
show this explicitly for $N_T=2$.
For smaller $\sigma$ in region III, as $\lambda$ decreases, the emergence of ``plateau''-like 
structures can be seen (see \Figref{delf-lambda-g1}). As will be shown, these flat areas correspond
to the model behaving as polynomial fits of different orders. These can be derived by
taking the limits $\sigma\to\infty$ and $\lambda\to0$ while maintaining $\sigma$ in a certain
proportion to $\lambda$.

\paragraph{$N_T=2:$}

In this case, the ML function is 
\ben
f\ML(x) = \alpha_1 \, e^{-x^2/2\sigma^2} + \alpha_2 \, e^{-(x-1)^2/2\sigma^2},
\een
and the weights are determined by solving
\ben
\Spvek{\alpha_1;\alpha_2} = \left( \begin{array}{ccc}
1+\lambda & e^{-1/2\sigma^2} \\ \\
e^{-1/2\sigma^2} & 1+\lambda \\
\end{array} \right)^{-1}
\Spvek{f_1;f_2},
\een
where $f_j = f(x_j)$.
The solution is
\bea
\alpha_1 &=& (f_1(1+\lambda) - e^{-1/2\sigma^2} f_2)/D, \\
\alpha_2 &=& (f_2(1+\lambda) - e^{-1/2\sigma^2} f_1)/D,
\eea
where $D = \det(K + \lambda I) = 1 + 2\lambda + \lambda^2 - e^{-1/2\sigma^2}$.
First, we expand in powers of $\sigma$ as $\sigma\to\infty$, 
keeping up to first order in $1/2\sigma^2$:
\bea
\alpha_1 &\approx& ((f_1 - f_2) + f_1 \lambda + f_2/2 \sigma^2)/D, \\
\alpha_2 &\approx& ((f_2 - f_1) + f_2 \lambda + f_1/2 \sigma^2)/D,
\eea
where
\ben
D \approx 2\lambda + \lambda^2 + 1/2\sigma^2.
\een
Finally
\ben
f\ML(x) \approx \bar\alpha + (\alpha_2 (2x-1) - \bar\alpha x^2)/2\sigma^2,
\een
where $\bar\alpha=\alpha_1 + \alpha_2$.
Next, we take $\lambda\to 0$. In this limit $D$ vanishes and the weights
diverge. The relative rate at which the limits are taken will affect the asymptotic
behavior of the weights. The form of $D$ suggests we take
\ben
\beta = \frac{1}{2\lambda\sigma^2},
\label{eq:beta}
\een
where $\beta$ is a constant. 

Taking $\sigma\to\infty$, we obtain a linear form:
\ben
f_{\beta}\ML(x) = \frac{\beta f_1 + \overline{f} + \beta (f_2 - f_1)x}{\beta+1},
\label{eq:fMLplateautransition}
\een
where $\overline{f}=\half(f_1 + f_2)$. The parameter $\beta$ smoothly connects the constant and linear
plateaus. When $\beta\to0$, we recover the constant form $f\ML(x) = \overline{f}$; when $\beta\to\infty$, we recover the linear form $f\ML(x) = f_1 + x(f_2 - f_1)$.

We can determine the shape of the transition between plateaus by substituting 
\Eqref{eq:fMLplateautransition} for $f\ML(x)$ into \Eqref{delf} for $\Delta f$. For simplicity's sake, we first define
\ben
h_{ij} = \int_0^1 dx\, \, x^i f^j(x), 
\een
since expressions of this form will show up in subsequent derivations in this work. Finally, we obtain
\bea
\label{eq:delf-asymptotic}
\Delta f_{\beta} &=& \frac{-2(\overline{f} + f_1\beta)h_{01}}{1+\beta} + \frac{2\beta(f_1 - f_2)h_{11}}{1+\beta} \nonumber \\
&&{} + \frac{(3 + 6\beta + 4\beta^2)\overline{f}^2 -f_1f_2\beta^2}{3(1+\beta^2)} + h_{02}.
\eea

\begin{figure}[tb]
\includegraphics[width=\figwidth]{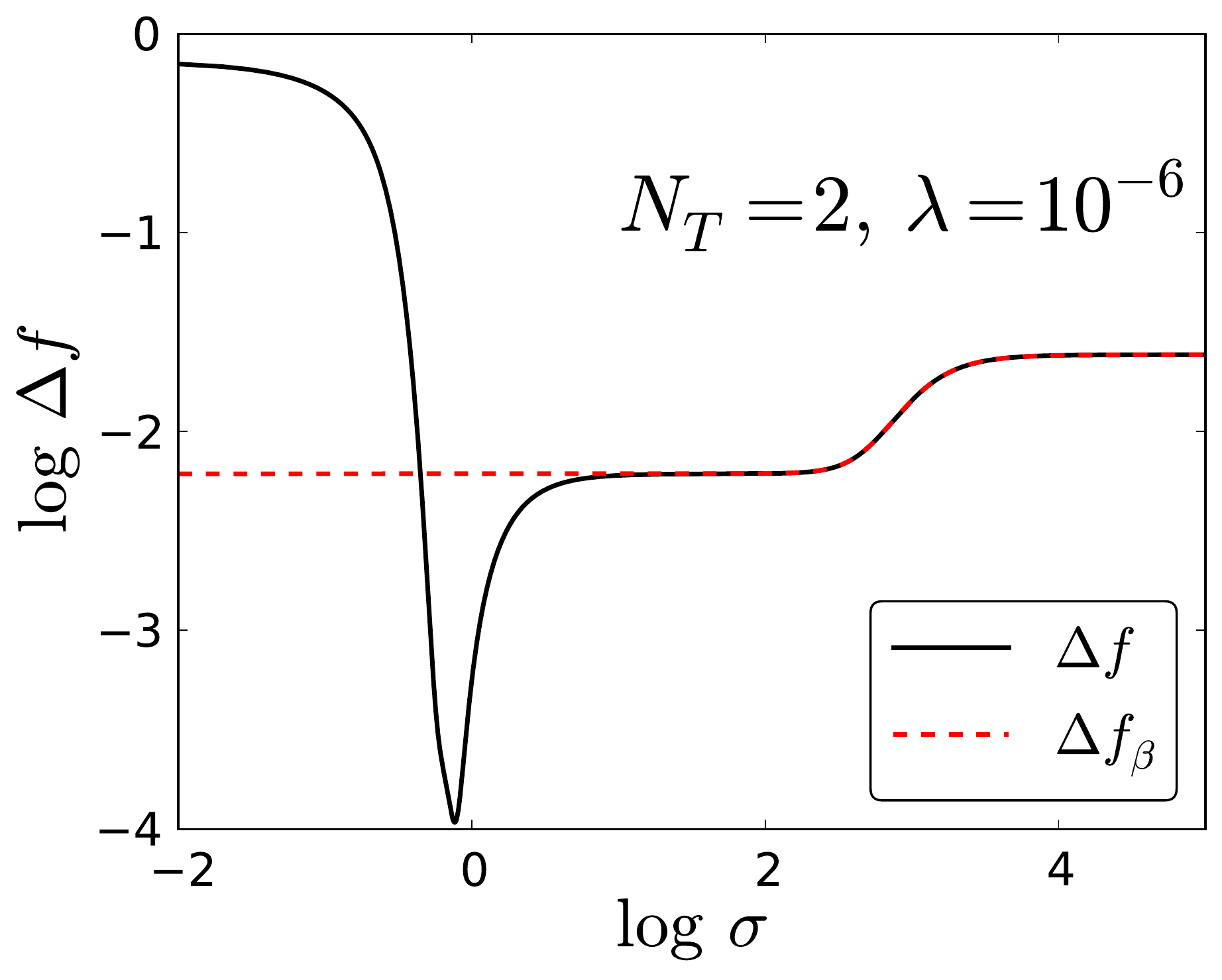}
\caption{Comparing the numerical $\Delta f$ (black solid), for $N_T=2$ and $\lambda=10^{-6}$, with
our asymptotic form $\Delta f_{\beta}$ (red dashed)
given by \Eqref{eq:delf-asymptotic}.
The asymptotic form accurately recovers the behavior of $\Delta f$ in the plateau regions, but
fails for small $\sigma$ as expected.}
\label{plateau-transition-NT2}
\end{figure}

In \Figref{plateau-transition-NT2}, we compare our numerical $\Delta f$
with the expansion \Eqref{eq:delf-asymptotic} showing the transition between the linear and constant plateaus.
In the case of $N_T=2$, only these two plateaus exist. In general, there will be at most $N_T$
plateaus, each corresponding to successively higher order polynomial fits. However, not all of these plateaus will necessarily emerge for a given
$N_T$; as we will show, the plateaus only become
apparent when $\lambda$ is sufficiently small, i.e., when the asymptotic behavior is reached, and when
$\sigma$ and $\lambda$ are proportional in a certain way similar to how we defined $\beta$.
This analysis reveals the origin of the plateaus. In the series expansion for $\sigma\to\infty$,
$\lambda\to 0$, certain terms corresponding to polynomial forms becomes dominant when
$\sigma$ and $\lambda$ remain proportional.

\paragraph{$N_T=3:$}

We proceed in the same manner for $N_T=3$, using $\beta$ and substituting into the analytical form of $f\ML(x)$ for this case to obtain an expression for $\Delta f_{\beta}$  (shown in the appendix). This expression is plotted in \Figref{plateau-transition-NT3}.

\begin{figure}[tb]
\includegraphics[width=\figwidth]{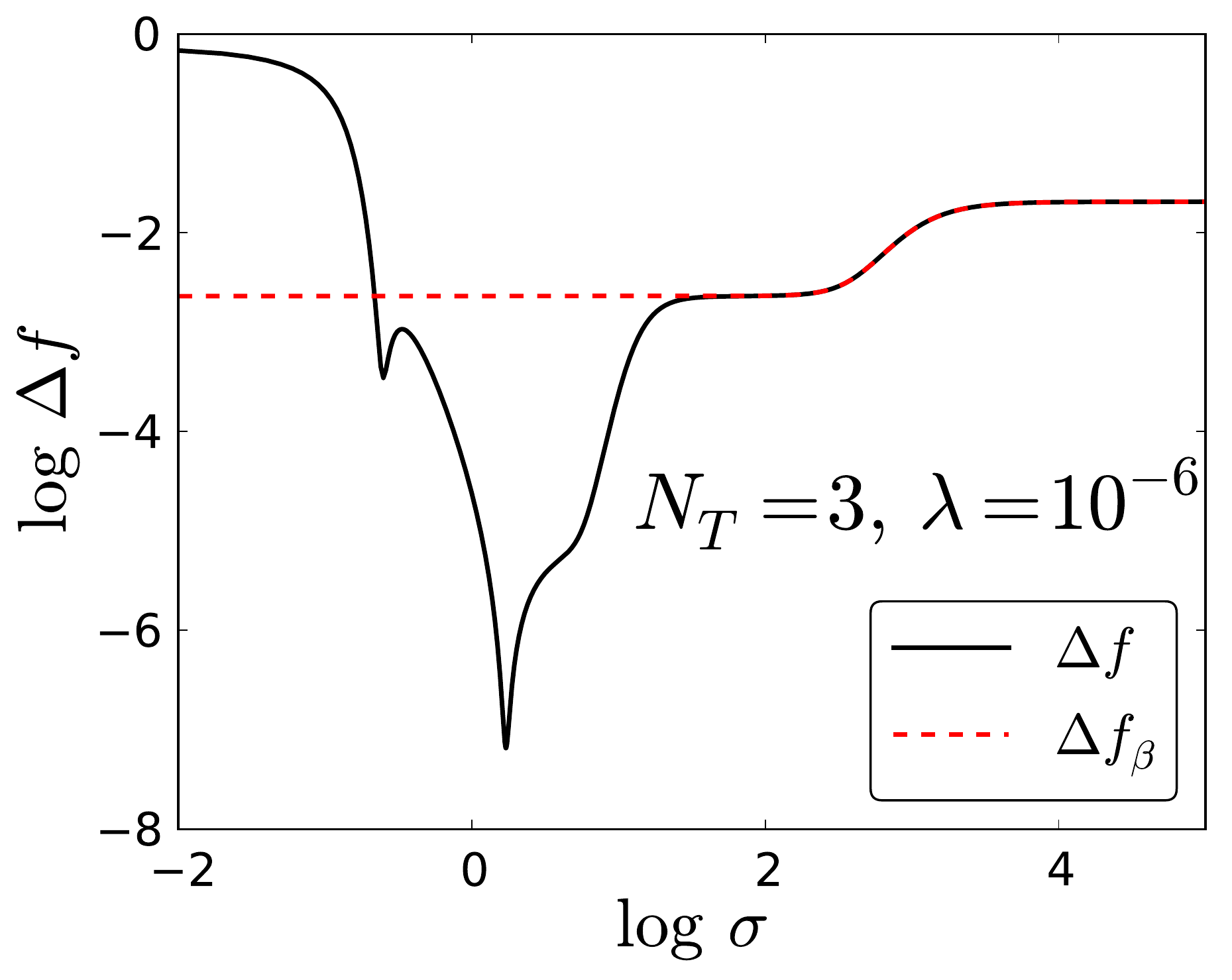}
\caption{Comparing the numerical $\Delta f$ (black solid), for $N_T=3$ and $\lambda=10^{-6}$, with
our asymptotic form $\Delta f_{\beta}$ (red dashed).
The asymptotic form accurately recovers the behavior of $\Delta f$ in the plateau regions, but
fails for small $\sigma$ as expected.}
\label{plateau-transition-NT3}
\end{figure}

To derive the limiting value of $\Delta f$ at each plateau for large 
$N_T$ and small $\lambda$,
we minimize the cost function \Eqref{costfunction} (which is equivalent to \Eqref{delf} in this limit),
assuming an $n$-th order polynomial form for $f\ML(x)$.
We define $c_n$ as the limiting value of $\Delta f$ for the $n$-th order plateau:
\ben
c_n = \lim_{\sigma\propto\lambda^{-1/2},N_T\to\infty} = \min_{\omega_i} \int_0^1 dx \, (f(x) - \sum_{i=0}^n \omega_ix^i)^2.
\label{delf-plateau-lim}
\een
For the constant plateau, $f\ML(x)$ assumes the constant form $a$; to minimize \Eqref{delf} with respect to $a$, we solve
\ben
\frac{d}{da} \int_0^1 dx \, (f(x) - a)^2 =0
\een
for $a$, obtaining 
\ben
a = \int_0^1 dx \, f(x),
\een
so that
\ben
f\ML(x) = h_{01}. 
\een
Thus, we obtain
\ben
c_0 = -h_{01}^2 + h_{02}. 
\label{plateau0}
\een
For our case with $f(x)=\cos(x)$, 
$c_0 = 0.0193$.

For the linear plateau, $f\ML(x)$ assumes the linear form $ax + b$; minimizing \Eqref{delf} with respect to $a$ and $b$, we find that 
\ben
f\ML(x) = (12\, h_{11} - 6\, h_{01})x + 4\, h_{01} - 6 \,h_{11},  
\een
yielding, via \Eqref{delf-plateau-lim}, 
\ben
c_1 = h_{02}-4 \left(h_{01}^2-3 \,h_{01} h_{11}+3\,
   h_{11}^2\right).
\label{plateau1}
\een
For our case with $f(x)=\cos(x)$,
$c_1 = 1\times 10^{-3}$.
The same procedure yields $c_2 = 2.25\times 10^{-6}$.

Next, we define 
\ben
\epsilon = \frac{1}{2\lambda^{1/2}\sigma^2}
\label{eq:alpha}
\een
as another parameter to relate $\sigma$ and $\lambda$. We choose to define this using the same motivation as for $\beta$, i.e., we examined our analytical expression for $f\ML(x)$ and picked this parameter to substitute in order for $\sigma$ and $\lambda$ to remain proportional in a specific way as they approach certain limits and to see what values $\Delta f$ takes for these limits (in particular, we are interested to see if we can obtain all 3 plateaus for $N_T=3$). In doing this, we obtain an expansion analogous to that of \Eqref{eq:delf-asymptotic} (shown as $\Delta f_{\epsilon}$ in the appendix). 

We plot this expression in \Figref{alpha-NT3}, alongside our numerical $\Delta f$ and the plateau limits, for $N_T=3$ and varying $\lambda$.
Note that the curves of the expansions are contingent on the value of $\lambda$; we do not retrieve all 3 plateaus for all of the expansions. Only the expansion curves corresponding to the smallest $\lambda$ ($10^{-10}$, the blue curve) and second smallest $\lambda$ ($10^{-6}$, the yellow curve) show broad, definitive ranges of $\sigma$ where they take the value of each of the 3 plateaus (for the dashed blue curve, this is evident for $c_1$ and $c_2$; the curve approaches $c_0$ for larger $\sigma$ ranges not shown in the figure), suggesting a specific proportion between $\sigma$ and $\lambda$ is needed for this to occur. For the solid numerical curves, only the blue curve manifests all 3 plateaus (like its expansion curve counterpart, it approaches $c_0$ for larger $\sigma$ ranges not shown); the other two do not obtain all 3 plateaus, regardless of the range of $\sigma$ (the solid red curve does not even go down as far as $c_2$). However, there appears to be a singularity for each of the expansion curves (the sharp spikes for the dashed curves) at certain values of $\sigma$ ($\epsilon$) depending on $\lambda$. This singularity emerges because our substitution of $\epsilon$ leads to an expression with $\epsilon$ in the denominator of our $\Delta f$ analytical form, which naturally has a singularity for certain values of $\epsilon$ depending on $\lambda$. 
\begin{figure}[tb]
\includegraphics[width=\figwidth]{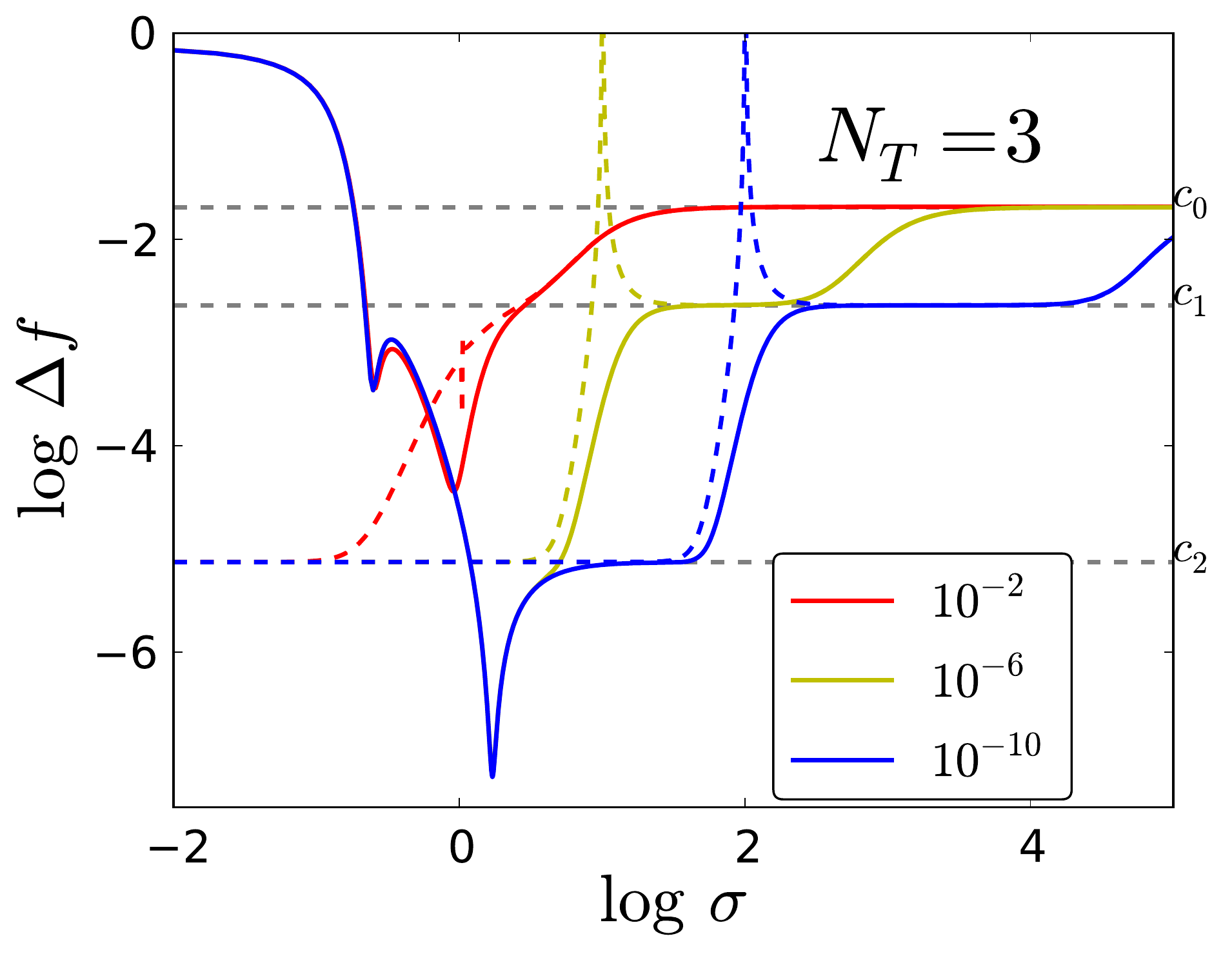}
\caption{The dependence of the model error on $\sigma$, for $N_T=3$ and varying $\lambda$. The solid curves are numerical; the dashed curves are expansions derived by using our expression for $\epsilon$ in \Eqref{eq:alpha} and substituting into $f\ML(x)$ in \Eqref{delf}. The legend gives the colors (for both the dashed and solid curves) corresponding to each $\lambda$.}
\label{alpha-NT3}
\end{figure}
Following the precedent set for $N_T=2$ and $N_T=3$, we can proceed in the same way for larger $N_T$ and perform the same analysis, where we expect to find higher order plateaus and the same behavior for limiting values of the parameters, including specific plateau values for $\Delta f$ when $\sigma$ and $\lambda$ are varied with respect to each other in certain ways analogous to that of the previous cases.
We would like to remark that plateau-like behaviors are well-known in statistical (online) learning in neural networks \cite{Saad98}. However, those plateaus are distinct from the plateau effects discussed here since they correspond to limits in the (online) learning behavior due to symmetries \cite{new2, new1} or singularities \cite{journals/nn/FukumizuA00, journals/nn/WeiA08} in the model.

\sssec{Length scale just right}

In the central region (see \Figref{regions}),
$\Delta f$ as a function of 
$\sigma$ has the shape of a valley. The optimum model, i.e.,
the model which gives the lowest error $\Delta f$, is found in
this region. 
\begin{figure}[tb]
\includegraphics[width=\figwidth]{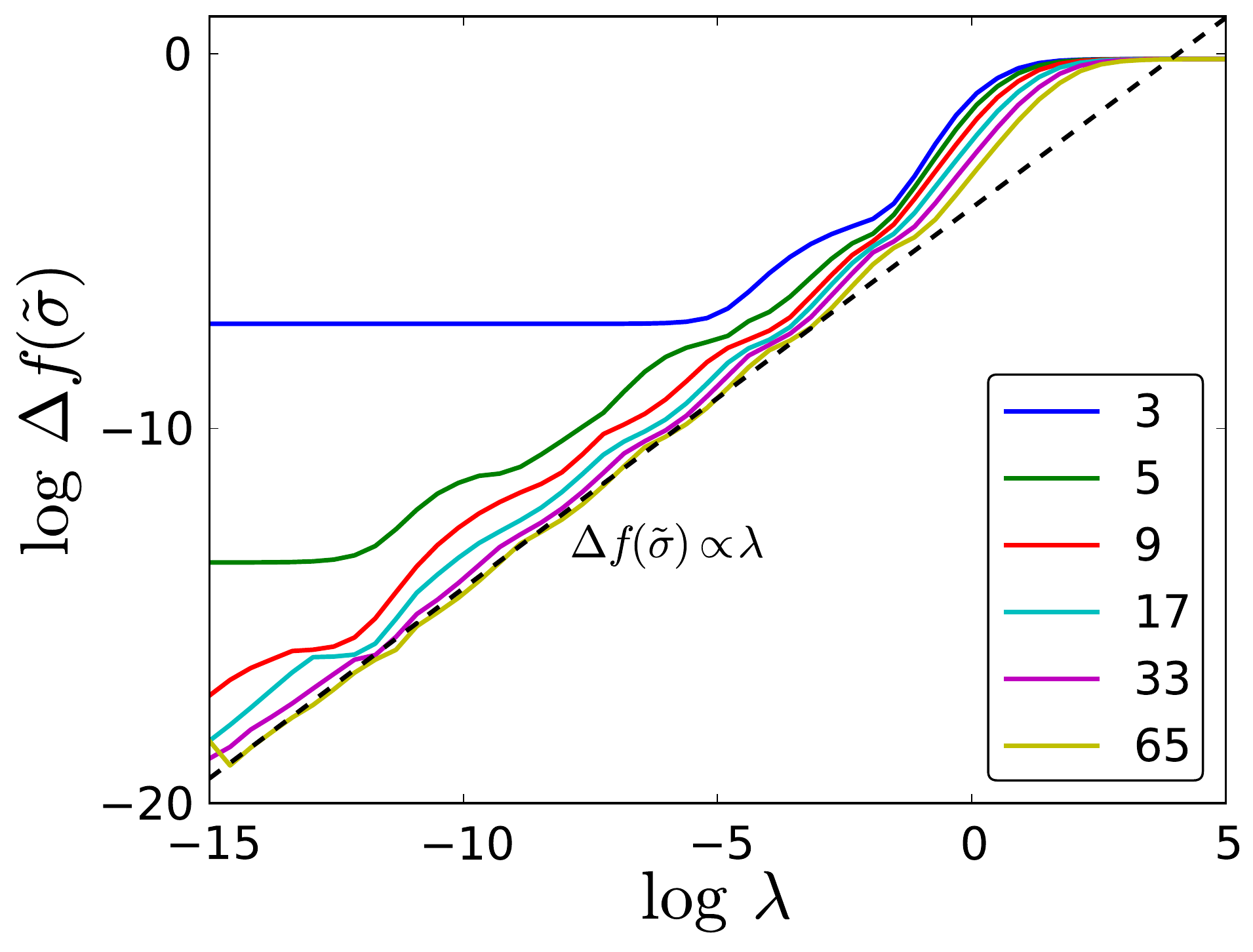}
\caption{The dependence of $\Delta f(\tilde \sigma)$ on $\lambda$ for various $N_T$. Here, $\tilde \sigma$ minimizes $\Delta f$ for fixed $N_T$ and $\lambda$. $N_T$ values for each curve are given in the legend. The dashed line shows a linear proportionality between $\Delta f(\tilde \sigma)$ and $\lambda$.}
\label{delfmin-vs-lambda-g1}
\end{figure}
For fixed $N_T$ and $\lambda$, we define the $\sigma$ that
gives the global minimum of $\Delta f$ as $\tilde \sigma$. In
\Figref{delfmin-vs-lambda-g1}, we plot the behavior of
$\Delta f(\tilde \sigma)$ as a function of $\lambda$.
Again, we observe three regions of different qualitative behavior.
For large $\lambda$, we over-regularize (as was shown in
\Figref{over-regularized}), giving the limiting value  
$\Delta f_0$ in \Eqref{limdelf-lambda-infty}. For moderate $\lambda$,
we observe an approximately linear proportionality between
$\Delta f(\tilde \sigma)$ and $\lambda$:
\ben
\Delta f(\tilde \sigma) \propto \lambda.
\label{delf-vs-lambda}
\een
However, for small enough $\lambda$,
there is vanishing regularization
\ben
\boldsymbol{\alpha}_\text{NF} = \lim_{\lambda \to 0} (\K + \lambda \I)^{-1} \f,
\een
yielding the {\em noise-free} limit of the model:
\ben
f_\text{NF}\ML(x) = \sum_{j=1}^{N_T} {\alpha_\text{NF}}_j k(x, x_j).
\een
In this case (for the Gaussian kernel), this limit exists
for all $\sigma$. The error of the noise-free model is
\ben
\Delta f_\text{NF} = \lim_{\lambda \to 0} \Delta f.
\een

%--------------------------------------------%
\ssec{Dependence on function scale}

We now introduce the parameter $\gamma$ into our simple one variable function, so that \Eqref{eq:model} becomes
\ben
f(x) = \cos(\gamma x).
\label{eq:model-gamma}
\een
For large values of $\gamma$, \Eqref{eq:model-gamma} becomes highly oscillatory; we extend our analysis here in order to observe the behavior of the model in this case.

\begin{figure}[tb]
\includegraphics[width=\figwidth]{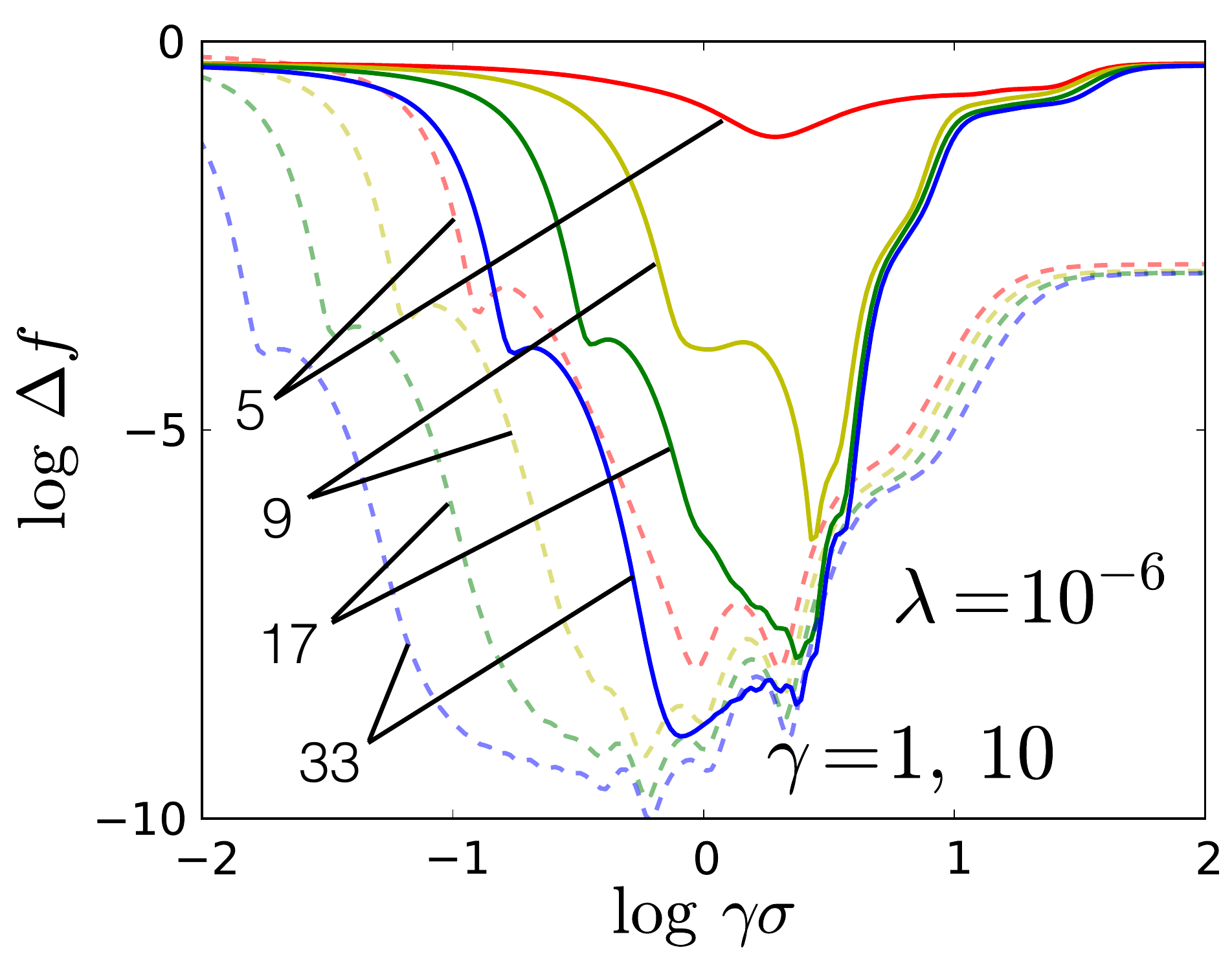}
\caption{Same as \Figref{delf-Nt}, except with $\sigma$ replaced by $\gamma\sigma$, where $\gamma=10$ for the solid curves and $\gamma=1$ for the dashed curves. The labels give the value of $N_T$ for each curve.}
\label{delf-Nt-gamma-g4}
\end{figure}

\Figref{delf-Nt-gamma-g4} shows $\Delta f$ as a function of $\gamma\sigma$ for various $N_T$ while fixing $\lambda=10^{-6}$ and $\gamma$=10 (solid lines), $\gamma=1$ (dashed lines). This is the same as that of \Figref{delf-Nt}, except with the additional $\gamma$ parameter. 
This figure demonstrates that the qualitative behaviors we observed in \Figref{delf-Nt} persist with the inclusion of the $\gamma$ parameter, complete with the characteristic ``valley'' shape emerging in the moderate $\sigma$ region for each $N_T$. 
Similarly, we see that $\Delta f$ decreases nearly monotonically for increasing $N_T$ for all $\gamma\sigma$, while opening up to the left as the Gaussians are better able to interpolate the function. The cusps, though not as pronounced, are still present to the left of the valleys, and their general shapes remain the same for increasing $N_T$.

\begin{figure}[tb]
\includegraphics[width=\figwidth]{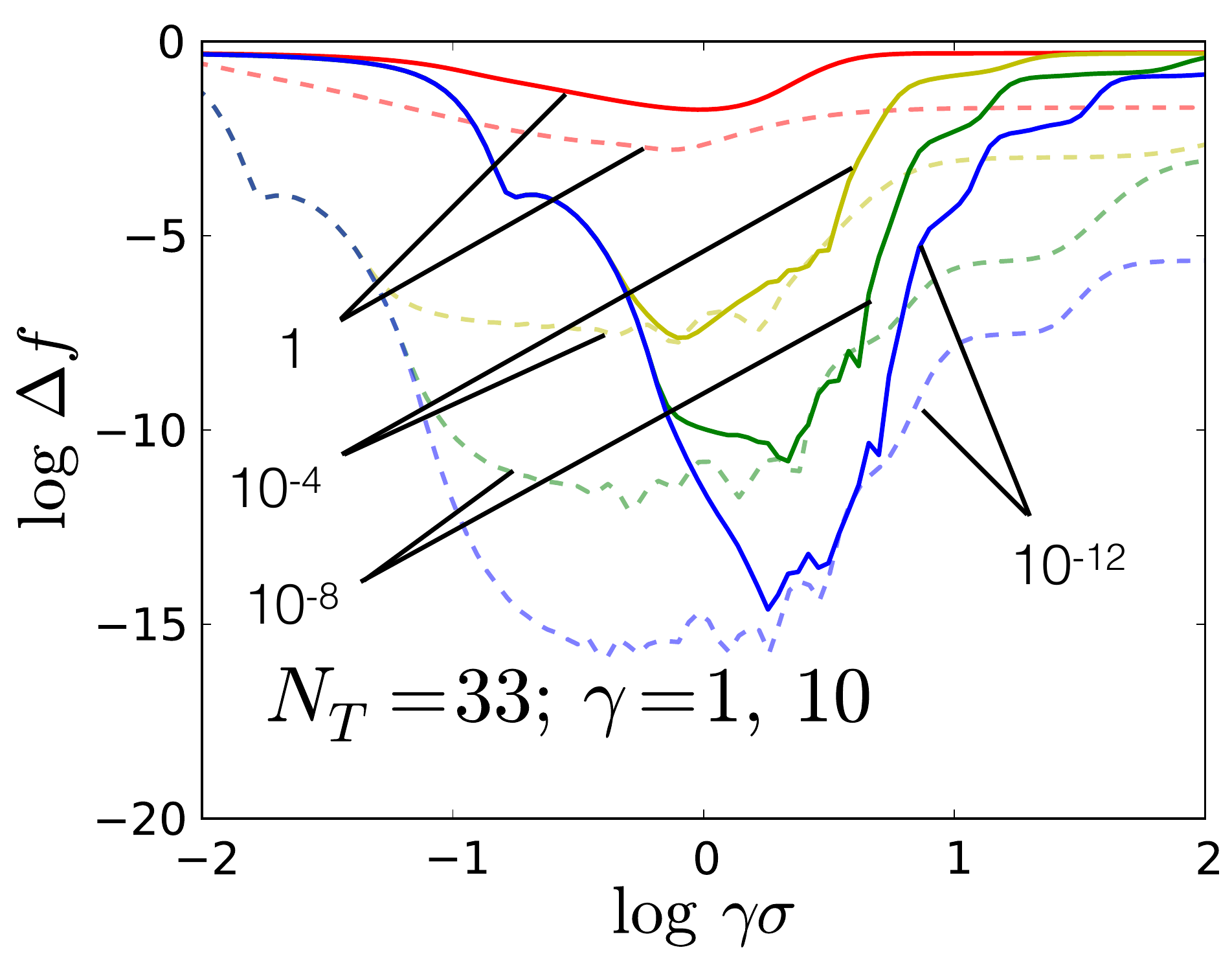}
\caption{Same as \Figref{delf-lambda-g1}, except with $\sigma$ replaced by $\gamma\sigma$, where $\gamma=10$ for the solid curves and $\gamma=1$ for the dashed curves. The labels give the value of $\lambda$ for each curve.}
\label{delf-lambda-g10}
\end{figure}

\Figref{delf-lambda-g10} shows $\Delta f$ as a function of $\gamma\sigma$ for various $\lambda$ while fixing $N_T=33$ and $\gamma$=10 (solid lines), $\gamma=1$ (dashed lines).  This is the same as \Figref{delf-lambda-g1}, except with the $\gamma$ parameter included. 
Like in \Figref{delf-lambda-g1}, as $\lambda$ decreases $\Delta f$ decreases nearly monotonically. The same qualitative features still hold, including the splitting-off of each lower-valued $\lambda$ curve further along $\sigma$.

\begin{figure}[tb]
\includegraphics[width=\figwidth]{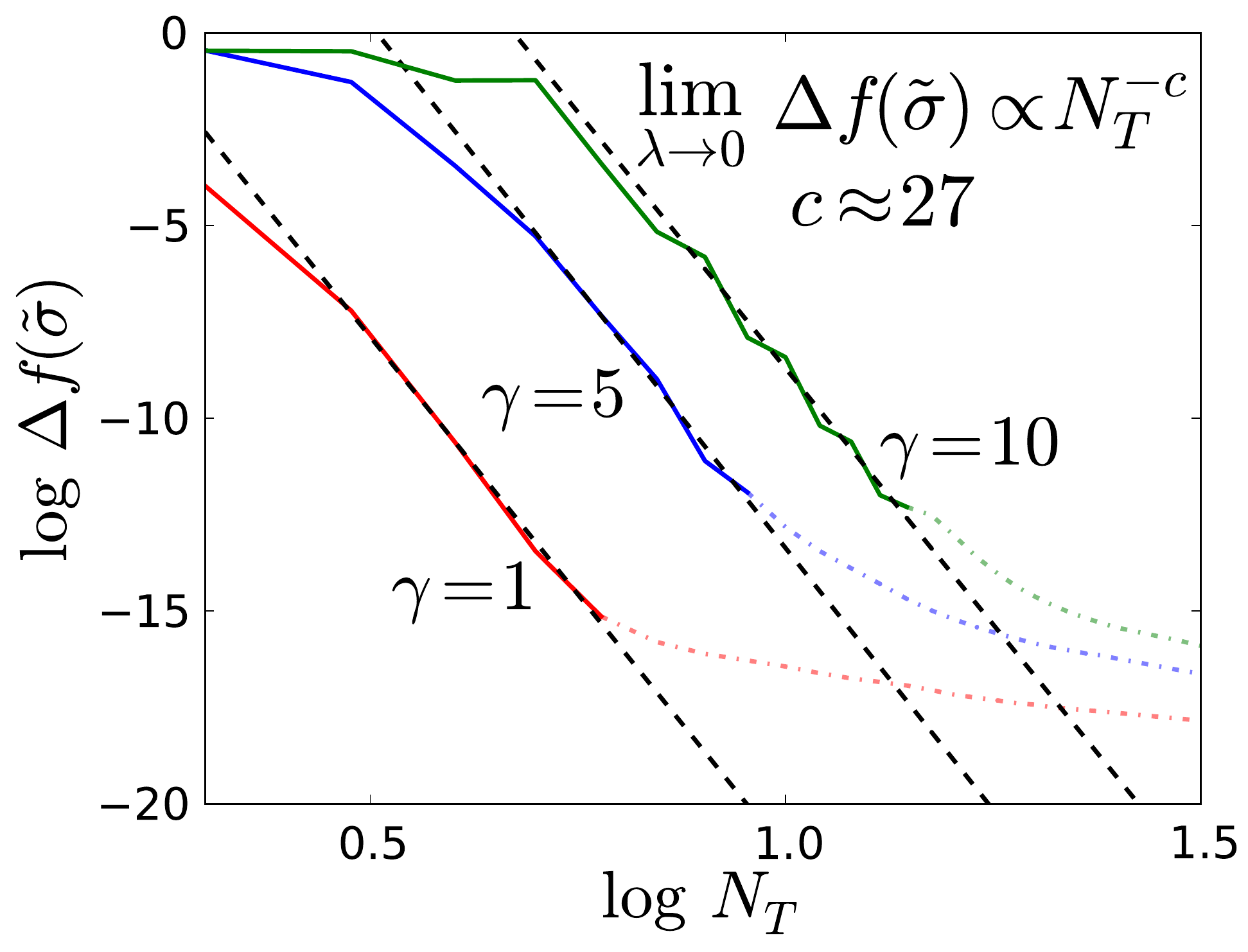}
\caption{The dependence of $\Delta f(\tilde \sigma)$ on $N_T$ for various $\gamma$. Here, $\tilde \sigma$ minimizes $\Delta f$ for fixed $N_T$ and $\lambda$. The solid portion of the line represents the limit at $\lambda \to$ 0 (the noise-free curve), while the dot-dashed continuation shows the decay for finite $\lambda$ ($\lambda = 10^{-14}$ is shown here). For large enough $N_T$ and $\lambda \to$ 0, $\Delta f$ has the asymptotic form given approximately by the linear fit here (dashed line). Note that, although this asymptotic form is independent of $\gamma$, for larger $\gamma$ the asymptotic region is reached at larger $N_T$.}
\label{delfmin-vs-Nt-noisefree}
\end{figure}

Next, we look at how the optimal model depends on $N_T$. In \Figref{delfmin-vs-Nt-noisefree}, we plot $\Delta f(\tilde \sigma)$ as a function of $N_T$, for various $\gamma$. For small $N_T$, there is little to no improvement in the model, depending on $\gamma$. For large $\gamma$, $f\ML(x)$ is rapidly varying and the model requires more training samples before it can begin to accurately fit the function. At this point, $\Delta f(\tilde \sigma)$ decreases as $N_{T}^{-c}$, where $c \approx 27$ is a constant independent of $\gamma$. This fast learning rate drops off considerably when $\Delta f$ is on the order of $\lambda$ (i.e., at the limit of machine precision), and $\Delta f(\tilde \sigma)$ levels off 
(as $\lambda$ corresponds to the leeway the model has for fitting training $f(x)$ values, i.e., to the
accuracy with which the model can resolve errors during fitting, it cannot improve the error much
beyond this value). In fact, it is known that the learning rate in the asymptotic limit is $1/N_T$ for faithful models (i.e., models that capture the structure of the data) and $1/\sqrt{N_T}$ for unfaithful models \cite{HTF09, muller1996numerical}. However, before the regularization kicks in $\Delta f$ is approximately the noise-free limit $\Delta f_\text{NF}$. If the noise-free limit were taken for all $N_T$, it appears that $\Delta f(\tilde \sigma)$ would decrease continually at the same learning rate:
\ben
\Delta f_\text{NF} \propto N_{T}^{-c}.
\label{eq:delf-hat-Nt}
\een

The learning rate here resembles the error decay of an integration rule, as our simple function is smooth and can always be approximated locally by a Taylor series expansion with enough points on the interval. However, the model here uses an expansion of Gaussian functions instead of polynomials of a particular order, and the error decays much faster than a standard integration rule such as Simpson's, which decays as $N_T^{-4}$ in the asymptotic limit. 
Additionally, \Eqref{eq:delf-hat-Nt} is independent of $\gamma$ since, for large enough $N_T$, the functions appear no more complex locally. The larger y-intercepts for the larger $\gamma$ curves in \Figref{delfmin-vs-Nt-noisefree} arise due to the larger number of points needed to reach this asymptotic regime, so the errors should be comparatively larger.

%--------------------------------------------%
\ssec{Cross-validation}\label{crossvalidation}

In previous works (\Ref{SRHM12,LSPH14}) applying ML to DFT, the hyperparameters of the model were optimized in order to find the best one, i.e., we needed to find the hyperparameters such that the error for the model is minimal on the entire test set, which has not been seen by the machine in training \cite{HMBF13}. We did this by using cross-validation, a technique whereby we minimize the error of the model with respect to the hyperparameters on a partitioned subset of the data we denote as the ${\textit validation\, set}$. Only after we have chosen the optimal hyperparameters through cross-validation do we test the accuracy of our model by determining the error on the test set.  We focus our attention on leave-one-out cross-validation, where the training set is randomly partitioned into $N_T$ bins of size one (each bin consisting of a distinct training sample). A validation set is formed by taking the first of these bins, while a training set is formed from the rest. The model is trained on the training set, and optimal hyperparameters are determined by minimizing the error on the singleton validation set. This procedure is repeated for each bin, so $N_T$ pairs of optimal hyperparameters are obtained in this manner; we take as our final optimal hyperparameters the median of each hyperparameter on the entire set of obtained hyperparameters.
The generalization error of the model with optimal hyperparameters will finally be tested on a test set, which is inaccessible to the machine in cross-validation.

\begin{figure}[tb]
\includegraphics[width=\figwidth]{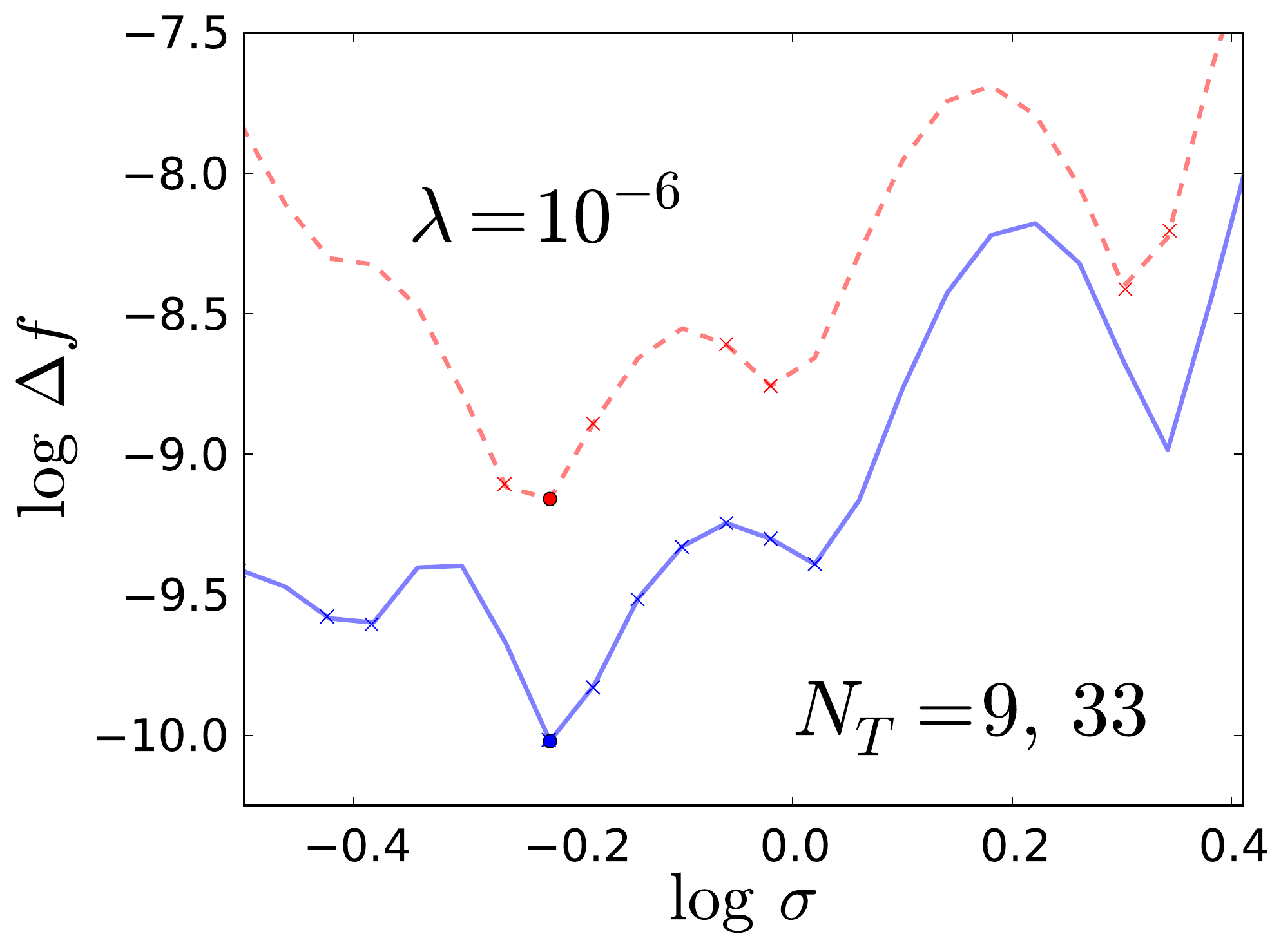}
\caption{The dependence of $\Delta f$ on $\sigma$, for $\lambda=10^{-6}$ and $N_T=9$ (curve shown in red with dashed lines) and $33$ (curve shown in blue with solid lines). The crosses denote $\Delta f$ for the optimized $\sigma$ values found from performing leave-one-out cross-validation (some of these are degenerate, so there are less than $N_T$ distinct crosses shown), while the dots denote $\Delta f(\tilde \sigma)$, the global minimum of $\Delta f$ over $\sigma$ (the crosses and dots are matched in color with the curves for each $N_T$).}
\label{crossvalidate_delf}
\end{figure}

Our previous works \cite{SRHM12,SRHB13,LSPH14} demonstrated the efficacy of cross-validation 
in producing an optimal model. Our aim here is to show how this procedure optimizes the model for our simple function on evenly-spaced training samples. We have thus far trained our model on evenly spaced points on
the interval $[0,1]$: $x_j = (j - 1)/(N_T - 1)$ for $j=1,\dots,N_T$. We want to compare how the model error determined in this way compares to the model errors using leave-one-out cross-validation to obtain optimal hyperparameters. In \Figref{crossvalidate_delf}, we plot the model error over a range of $\sigma$ values (we fix $\lambda=10^{-6}$ and we use $N_T=9$ and $N_T=33$; compare this with \Figref{delf-Nt}). For each $N_T$, we perform leave-one-out cross-validation (using our fixed $\lambda$ so that we obtain optimal $\sigma$), yielding $N_T$ optimal $\sigma$ values; we plot the model errors for each of these $\sigma$. We also include the global minimum error $\Delta f(\tilde \sigma)$ for each $N_T$ to show how they compare to the errors for the optimal $\sigma$. Looking at \Figref{crossvalidate_delf}, we see that the optimal $\sigma$ values all yield errors near $\Delta f(\tilde \sigma)$ and within the characteristic ``valley" region, demonstrating that leave-one-out cross-validation indeed optimizes our model. With this close proximity in error values established, we can thus reasonably estimate the error of the model for the optimal $\sigma$ (for a given $\lambda$) by using $\tilde \sigma$.

%Our previous works demonstrated the efficacy of cross-validation in producing a robust model. Our aim here is to show that this robustness persists for cross-validation on our present model, which we have thus far only treated with evenly-spaced training samples. We show this robustness by comparing the error of our model using optimal hyperparameters with that of the error determined on an evenly-spaced training set (which is simply the $\Delta f$ that we have been using throughout this paper).
%Once this robustness has been established, we will be able to reasonably estimate the error of the model for the optimal $\sigma$ (for a given $\lambda$) by using $\tilde \sigma$, which can expedite problems that would otherwise be too costly to be feasible. 
%We plot $\Delta f$ vs $\sigma$ alongside $\Delta f$ for each optimal $\sigma$ and $\Delta f(\tilde \sigma)$ in \Figref{crossvalidate_delf}, where we see that the optimized $\sigma$ all yield $\Delta f$ near $\Delta f(\tilde \sigma)$.

%--------------------------------------------%

%--------------------------------------------%

\sec{Application to density functionals}
\label{dft-section}

A canonical quantum system used frequently to explore basic quantum principles and as a proving ground for approximate quantum methods is the particle in a box. In this case, we confine one fermion to a 1d box with hard walls at $x=0,1$, with the addition of the external potential $v(x)$ in the interval $x\in[0,1]$. The equation that governs the quantum mechanics is the familiar one-body Schr{\"o}dinger equation
\ben
\left( -\half \frac{\partial^2}{\partial x^2} + v(x) \right)\phi(x) = \epsilon \phi(x).
\label{schroeq}
\een
A solution of this equation gives the orbitals $\phi_j(x)$ and energies $\epsilon_j$. For one fermion, only the lowest energy level is occupied. The total energy is $E=\epsilon_1$, the potential energy is
\ben
V = \int dx\, n(x) v(x)
\een
(where $n(x)=|\phi(x)|^2$ is the electron density),
and the KE is $T = E - V$.
In the case of one particle, the KE can be expressed exactly in terms of the electron density, 
known as the von Weizs{\"a}cker functional \cite{W35}
\ben
T\W = \int dx\, \frac{n'(x)^2}{8n(x)},
\een
where $n'(x) = dn/dx$. Our goal here in this section is not to demonstrate the efficacy 
of ML approximations for the KE in DFT (which is the subject of other works \cite{SRHM12,SRHB13}), 
but rather to study the properties of the ML approximations with respect to those applications.

We choose a simple potential inside the box,
\ben
v(x) = -D \sin^2{\pi x} ,
\een
to model a well of depth $D$, which has also been used in the study of semiclassical methods \cite{CLEB11}.
To generate reference data for ML to learn from, we solve \Eqref{schroeq} numerically by discretizing space onto a uniform grid, $x_j = (j-1)/(N_G-1)$, for $j=1,\dots,N_G$, where $N_G$ is the number of grid points. Numerov's method is used to solve for the lowest energy orbital and its corresponding eigenvalue.  We compute $T$ and $n(x)$, which is represented by its values on the grid. For a desired number of training samples $N_T$, we vary $D$ uniformly over the range $[0,100]$, inclusive, generating $N_T$ pairs of electron densities and exact KEs. Additionally, a test set with $500$ pairs of electron densities and exact KEs is generated.

As in the previous sections, we use KRR to learn the KE of this model system. The formulation is identical to that of \Ref{SRHM12}:
\ben
T\ML[\n] = \sum_{j=1}^{N_T} \alpha_j k[\n, \n_j],
\een
where $k$ is the Gaussian kernel
\ben
k[\n, \n'] = \exp(-\| n - n' \|^2/(2\sigma^2)),
\een
and
\ben
\| n - n' \|^2 = \Delta x \sum_{j=1}^{N_G} (n(x_j) - n'(x_j))^2,
\een
where $\Delta x=1/(N_G-1)$ is the grid spacing.
The weights are again given by \Eqref{weights}, found by minimizing the cost 
function in \Eqref{costfunction}.

In analogy to \Eqref{delf}, we measure the error of the model as the total squared error integrated
over the interpolation region
\ben
\Delta T = \int_{0}^{100} dD\, (T\ML[n_D] - T[n_D])^2,
\een
where $n_D$ is the exact density for the potential with well depth $D$, and $T[n_D]$ is the exact corresponding KE. We approximate the integral by Simpson's rule evaluated on $D$ sampled over the test set (i.e., 500 values spaced uniformly over the interpolation region). This sampling is sufficiently dense over the interval to give an accurate approximation to $\Delta T$.

\begin{figure}[tb]
\includegraphics[width=\figwidth]{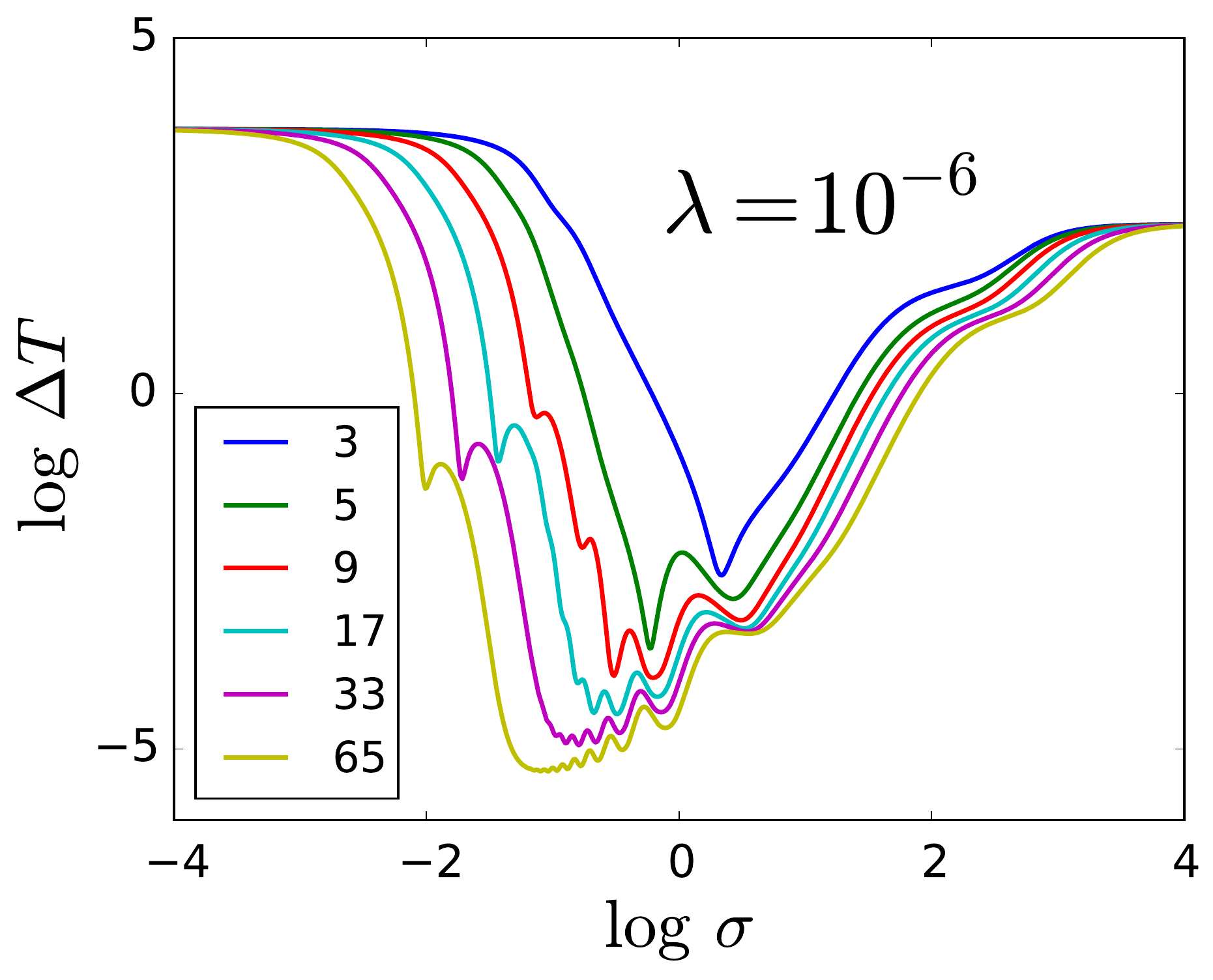}
\caption{The error of the model, $\Delta T$ (Hartree), as a function of $\sigma$,
for fixed $\lambda=10^{-6}$. $N_T$ values for each curve are given in the legend.}
\label{box-delT-NT}
\end{figure}

In \Figref{box-delT-NT}, we plot $\Delta T$ as a function of the length scale of the Gaussian kernel, $\sigma$, for various training set sizes $N_T$. Clearly, the trends are very similar to \Figref{delf-Nt}: the transition $\sigma_s$ between the regions I and II becomes smaller as $N_T$ increases, the valley in region II widens, and region III on the right remains largely unchanged. The dependence of $\sigma_s$ on $N_T$ appears to follow the same power law $\sigma_s \propto N_T^p$, but the value of $p$ is different in this case. A rough estimate yields $p\approx -0.8$, which is similar to $p=-1$ for the simple cosine function explored in the previous sections, but the details will depend on the specifics of the data.

\begin{figure}[tb]
\includegraphics[width=\figwidth]{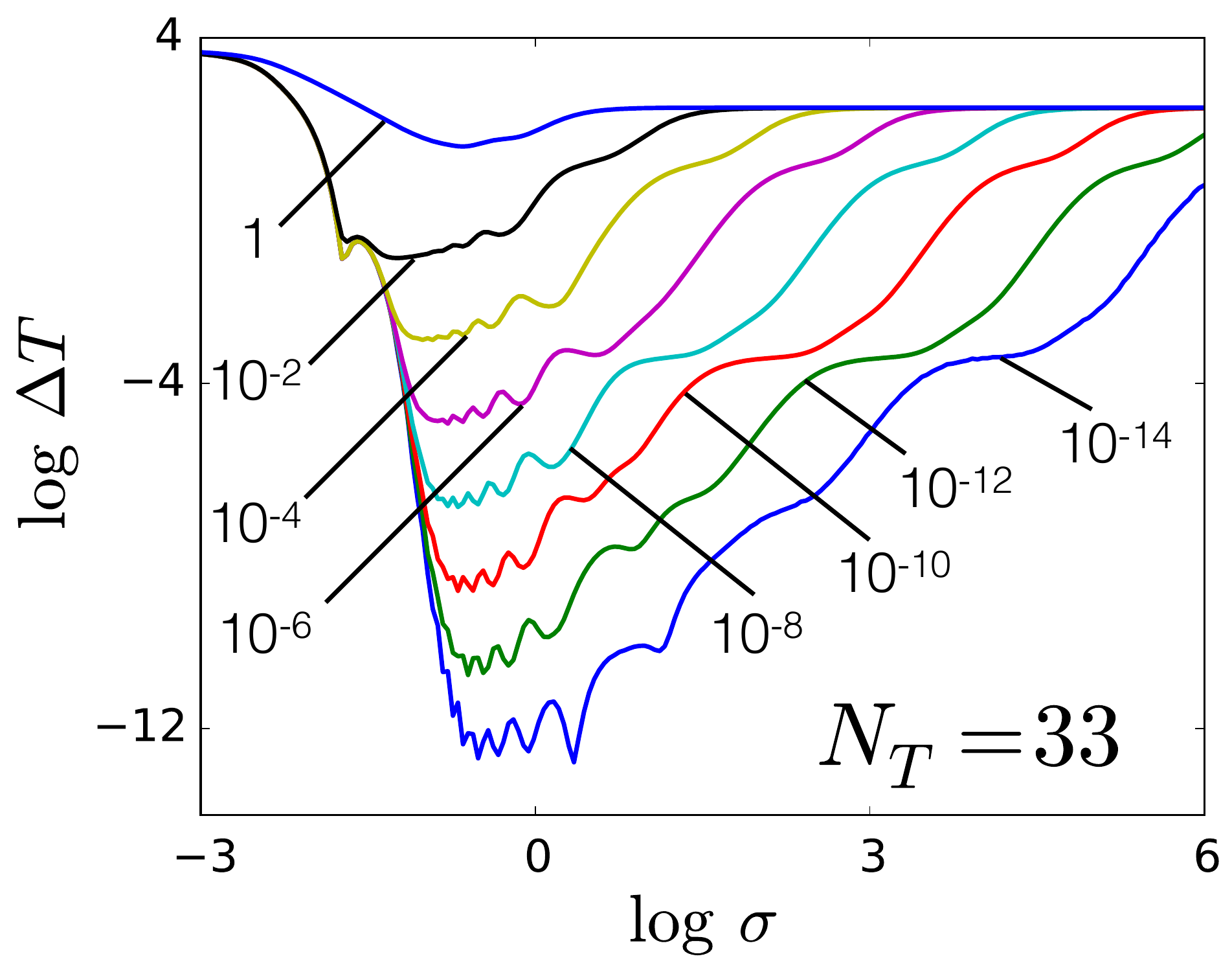}
\caption{The error of the model, $\Delta T$ (Hartree), as a function of $\sigma$, for various $\lambda$ with $N_T=33$.
The labels give the value of $\lambda$ for each curve.}
\label{box-delT-nl}
\end{figure}

Similarly, \Figref{box-delT-nl} shows the same plot but with $N_T$ fixed and $\lambda$ varied. Again, the same features are present as in \Figref{delf-lambda-g1}, i.e., three regions with different qualitative behaviors. In region I, $\Delta T$ has the same decay shape as the kernel functions (Gaussians) begin to overlap significantly, making it possible for the regression to function properly and fit the data. For large values of the regularization strength $\lambda$, the model over-regularizes, yielding high errors for any value of $\sigma$. As $\lambda$ decreases, the weights are given more flexibility to conform to the shape of KE functional. 
Using the same definition for the estimation of $\sigma_s$ in \Eqref{lsL}, the median nearest neighbor distance over this training set gives $\sigma_s=0.019$. We then have $\log \sigma_s=-1.72$, which matches the boundary between regions I and II in \Figref{box-delT-nl}.
In region III, the same plateau features emerge for small $\lambda$. Again, these plateaus occur when polynomial forms of the regression model become dominant for certain combinations of the parameters $\sigma$, $\lambda$, and $N_T$.

From \Eqref{limdelf-sigma-0} and \Eqref{limdelf-lambda-infty}, we showed that the model error will tend to the benchmark error while $\sigma\to 0$ or $\lambda\to \infty$.
Similarly to \Eqref{delf0}, we can also define the benchmark error when $T\ML[\n]\equiv 0$ for this data set as
\ben
\Delta T_0=\int_0^{100} dD \, \, T^2[\n_D].
\label{delT0}
\een
Evaluating the above integral numerically on the test set, we obtain $\log \Delta T_0=3.7$. This matches the error when $\sigma\to 0$ in \Figref{box-delT-NT} and \Figref{box-delT-nl}.

\begin{figure}[tb]
\includegraphics[width=\figwidth]{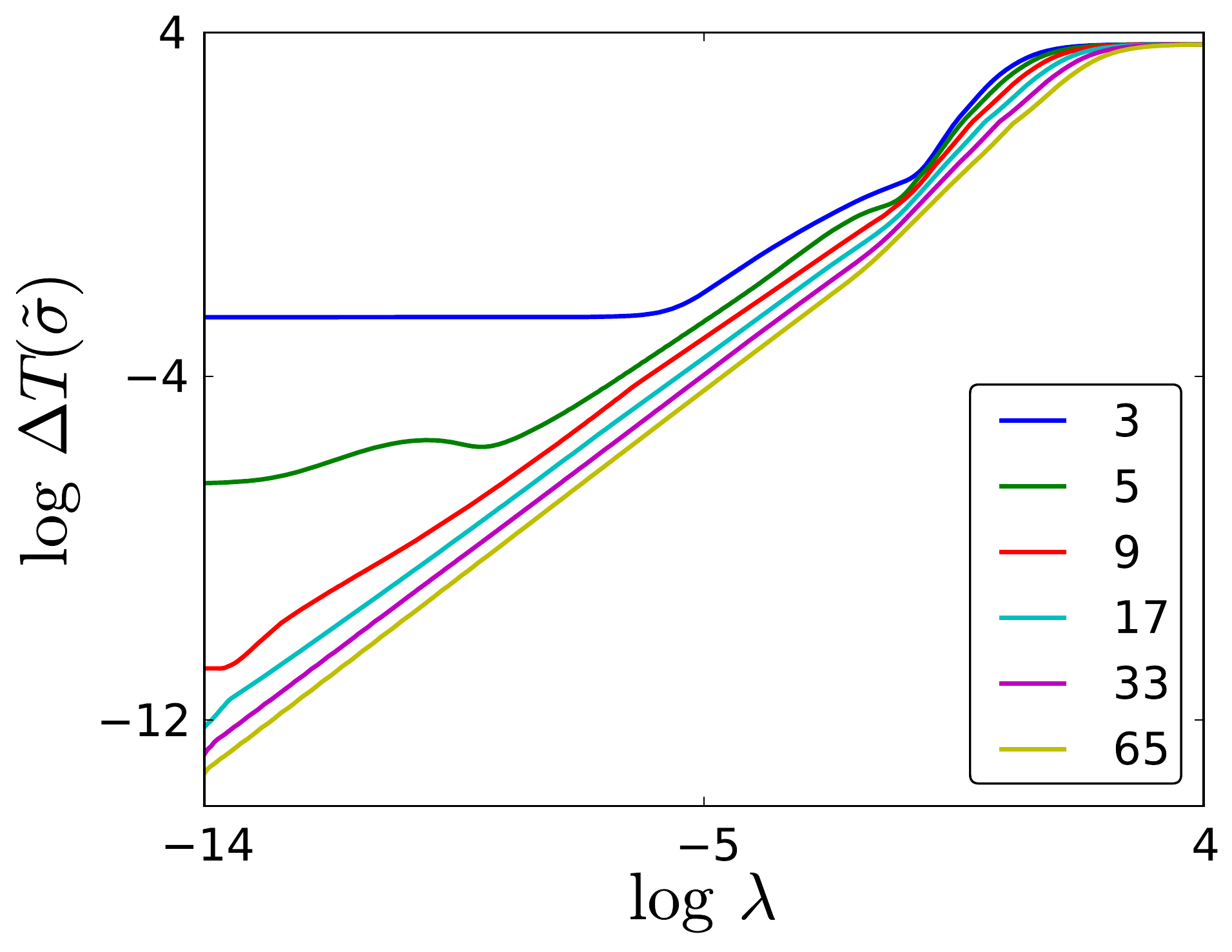}
\caption{The dependence of $\Delta T(\tilde \sigma)$ on $\lambda$ for various $N_T$. Here, $\tilde \sigma$ minimizes $\Delta f$ for fixed $N_T$ and $\lambda$. $N_T$ values for each curve are given in the legend.}
\label{delT-vs-lambda}
\end{figure}

We define the $\sigma$ that gives the global minimum of $\Delta T$ as $\tilde \sigma$; similarly to \Figref{delfmin-vs-lambda-g1}, we plot the optimal model error $\Delta T(\tilde \sigma)$ as a function of $\lambda$ in \Figref{delT-vs-lambda}.
For large $\lambda$, we overregularize the model; the model error tends to the benchmark error in \Eqref{delT0}. For moderate $\lambda$, we observe the same linear proportionality $\Delta T(\tilde \sigma)\propto \lambda$ as in \Figref{delfmin-vs-lambda-g1}.

In this toy system, the prediction of the KE from KRR models shares properties similar to those that we explored in learning the 1d cosine function. Now we will consider up to 4 noninteracting spinless fermions under a potential with 9 parameters as reported in~\Ref{SRHM12}. 
\ben
v(x) = -\sum_{i=1}^3 a_i \exp\left[-(x-b_i)^2/(2 c_i^2)\right].
\label{9-vp}
\een
These densities are represented on $N_G=500$ evenly spaced grid points in $0\leq x\leq1$.
Here a model is built using $N_T/4$ pairs of electron densities and exact KEs for each particle number $N=1,2,3,4$, respectively. Thus, the size of the training set is $N_T$. 1000 pairs of electron densities and exact KEs are generated for each $N$, so the size of the test set is $S=4000$. Since there are 9 parameters in the potential, we cannot define the error as an integral, so we use summation instead. Thus, the error on the test set is defined as the mean square error (MSE) on the test densities
\ben
\Delta T = \sum_{j=1}^{S} (T\ML[n_j] - T[n_j])^2/S.
\label{9-error}
\een

\begin{figure}[tb]
\includegraphics[width=\figwidth]{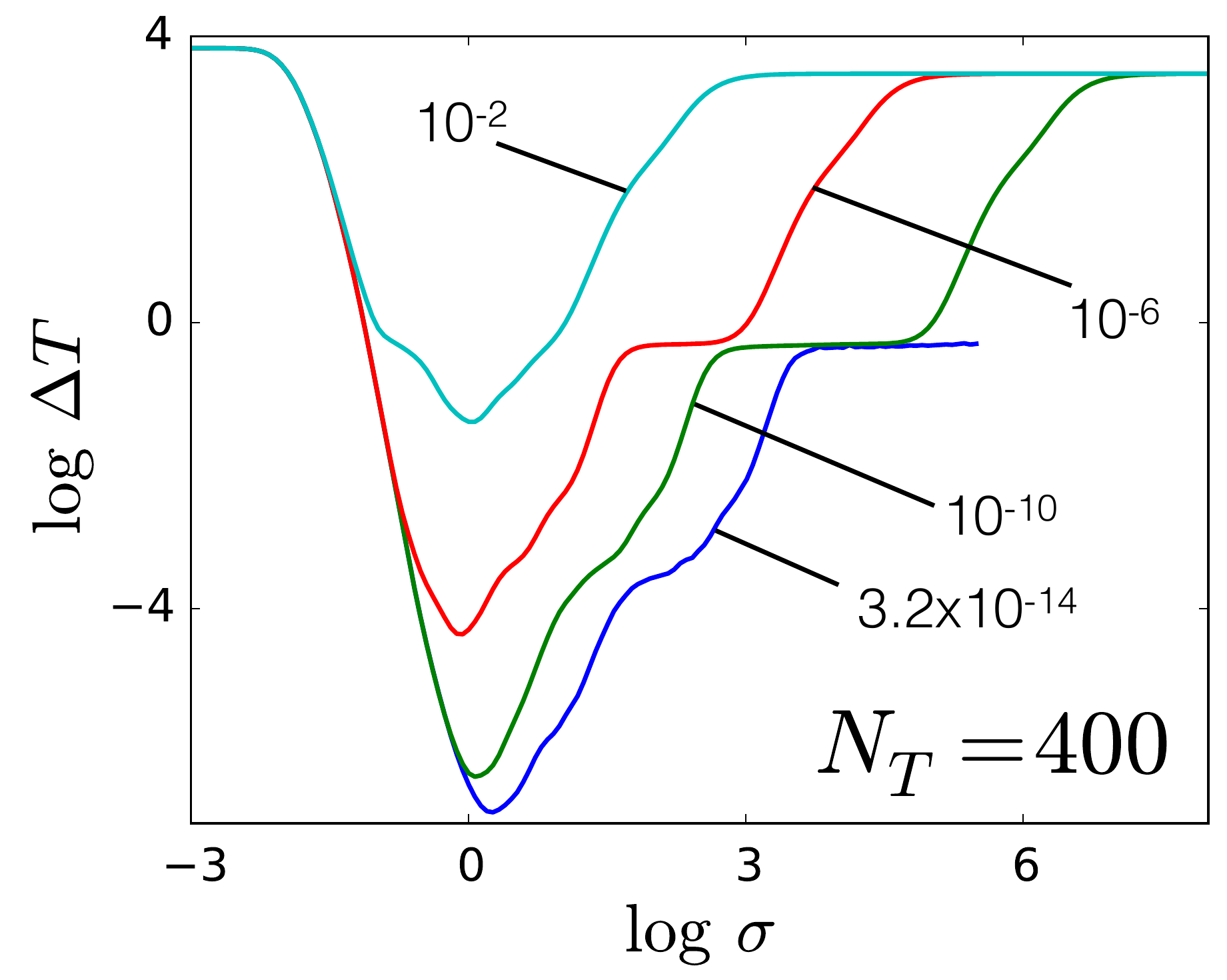}
\caption{Dependence of the model error (as a function
of $\sigma$) for various $\lambda$. The labels give the value of $\lambda$ for each curve. The $\lambda=3.2\times 10^{-14}$ curve is not plotted for $\log \sigma>5.5$ due to the numerical instability that occurs when $\lambda$ is small for large $\sigma$.}
\label{9-box-delT-nl}
\end{figure}

\Figref{9-box-delT-NT} shows the error of the model as a function of $\sigma$ with various $N_T$ for fixed $\lambda=10^{-10}$. Although this system is more complicated than the previous two systems discussed in this paper, the qualitative behaviors in \Figref{9-box-delT-NT} are similar to \Figref{delf-Nt} and \Figref{box-delT-NT}.
Table I in~\Ref{SRHM12} only shows the model error with optimized hyperparameters for $N_T=400$. In \Figref{9-box-delT-nl}, model errors varying with a wide range of $\sigma$ values are shown for 4 different values of $\lambda$ \footnote{For large $\sigma$, if $\lambda$ is small, there will be numerical instability in the computation of $(\K + \lambda\I)^{-1}$. Thus, the $\lambda=3.2\times 10^{-14}$ curve is not plotted for $\log \sigma>5.5$.}. The qualitative properties in \Figref{9-box-delT-nl} are similar to \Figref{delf-lambda-g1} and \Figref{box-delT-nl}. For example, the existence of three regions with distinctly different behavior for the model error can be ascertained just like before. In region I, error curves with different $\lambda$ will all tend to the same benchmark error limit when $\sigma\to 0$. 
The median nearest neighbor distance over this training set gives $\sigma_s=0.022$.
In \Figref{9-box-delT-nl}, the boundary between region I and region II is well estimated by $\log \sigma_s=-1.66$.
In region III, the familiar plateau features emerge. In region II, where $\sigma$ is such that the model is optimal or close to it, we find that the model with hyperparameters $\sigma=1.86,\, \lambda=3.2\times10^{-14}$ performs the best. The MSE for this model is $1.43\times10^{-7}$ Hartree. Another common measure of error is the mean absolute error (MAE), which is also used in~\Ref{SRHM12}. The MAE of this model is $1.99\times10^{-4}\,\text{Hartree}=0.12 \,\text{kcal}/\text{mol}$. This result is consistent with the model performance reported in~\Ref{SRHM12}\footnote{
In~\Ref{SRHM12}, the densities were treated as vectors. Here, we treated the densities as functions, so the length scale mentioned here is related to the length scale in~\Ref{SRHM12} by a factor of $\sqrt{\Delta x}=0.045$.}.

%--------------------------------------------%

\sec{Conclusion}

In this work, we have analyzed the properties of KRR models with a Gaussian kernel
applied to fitting a simple 1d function.
In particular, we have explored regimes of distinct qualitative behavior and derived
the asymptotic behavior in certain limits.
Finally, we generalized our findings to the problem of learning the KE functional of a
single particle confined to a box and subject to a well potential with variable depth.
Considering the vast difference in nature of the two problems compared in this work, a 1d cosine function and the KE as a functional of the electron density (a very high-dimensional object), 
the similarities of the measures of error $\Delta f$ and $\Delta T$ between each other are remarkable.
This analysis demonstrates that much of the behavior of the model can be rationalized by 
and distilled down to the properties of the kernel.
Our goal in this work was to deepen our understanding of how the performance
of KRR depends on the parameters qualitatively, in particular in the case that is relevant for MLA in DFT, namely the one of noise-free
data and high-dimensional inputs, and how one may determine {\em a-priori} which
regimes the model lies in. From the ML perspective the scenario analyzed in this work was an unusual one: small data, virtually no noise, low dimensions and high complexity. The effects found are not only interesting from the physics perspective, but are also illuminating from a learning theory point of view. However, in ML practice the extremes that contain plateaus or the ``comb" region will not be observable, as the practical data with its noisy manifold structure will confine learning in the favorable region II.   Future work will focus on theory and practice in order to improve learning techniques for low noise problems.

%--------------------------------------------%

\acknowledgments

The authors would like to thank
NSF Grant No. CHE-1240252 (JS, LL, KB), the Alexander von Humboldt Foundation (JS),
the
Undergraduate Research
Opportunities Program (UROP)
and the Summer Undergraduate Research
Program (SURP) (KV) at UC Irvine for funding.
MR thanks O. Anatole von Lilienfeld for support via SNSF Grant No. PPOOP2 138932.

%--------------------------------------------%

\appendix
\section{Expansion of $\Delta f$ for $\sigma \ll \sigma_s$}
\bea
\Delta f &=& \int_0^1 dx\, f(x)^2 - 2\sum_{j=1}^{N_T} \alpha_j \int_0^1 dx\, f(x) k(x, x_j) \nonumber \\
&& {} + \sum_{i,j=1}^{N_T} \alpha_i \alpha_j \int_0^1 dx\, k(x, x_i) k(x, x_j).
\label{delf-sigma-ll-sigmaL}
\eea
where the first integral is given in \Eqref{delf0cos}. Next
\ben
\int_0^1 dx\, f(x) k(x, x_j) = \sqrt{\frac{\pi}{8}}\sigma e^{-(\gamma \sigma)^2/2} ( C_j + C_j^* ),
\een
where
\ben
C_j = e^{\i \gamma x_j} \left( \erf\left(\frac{x_j - \i \gamma \sigma^2}{\sigma\sqrt{2}}\right) + \erf\left(\frac{1-x_j + \i \gamma \sigma^2}{\sigma\sqrt{2}}\right) \right),
\een
$\erf$ is the error function, and $C^*$ denotes the complex conjugate of $C$. The last integral is
\bea
\int_0^1 dx\, k(x, x_i) k(x, x_j) &=& \frac{\sigma\sqrt{\pi}}{2} e^{-(x_i - x_j)^2/(4\sigma^2)} \nonumber \\
&& \hspace{-3.5cm}\times \left(
\erf\left(\frac{x_i + x_j}{2\sigma}\right) -
\erf\left(\frac{x_i + x_j - 2}{2\sigma}\right)
\right).
\eea

\section{$\Delta f_{\beta}$ for $N_T=3$}
\bea
\Delta f_{\beta} &=& h_{01} \left(\frac{1}{3} (-5 f_1-2
   f_2+f_3)+\frac{f_1-f_3}{\beta +1}\right) \nonumber \\
&&{} + \frac{2 \beta \, h_{11} (f_1-f_3)}{\beta +1} + h_{02} + C,
\eea
where 
\bea
C &=& (\beta ^2 (7 f_1^2+2 f_1 (4 f_2+f_3)+4
   f_2^2+8 f_2 f_3 \nonumber \\
&&{} +7 f_3^2)+36 (2 \beta +1)
   \overline{f}^2)/(36 (\beta +1)^2),
\eea 
and where $\overline{f}=\frac{1}{3}(f_1+f_2+f_3)$.

\section{$\Delta f_{\epsilon}$ for $N_T=3$}
\bea
\Delta f_{\epsilon} &=& C_1\,h_{01} + C_2\,h_{11}  + C_3\, h_{21} + h_{02} + C_4,
\eea
where
\bea
C_1 &=& ((\sqrt{\lambda } (48 \overline{f}-\epsilon ^2
   (17 f_1+4 f_2+f_3))\nonumber \\
&&{}-2 \epsilon  (f_1
   (\epsilon ^2-20)-8 f_2+4 f_3)-4 \lambda
   \epsilon  (f_2+4 f_3)))/\nonumber \\
&&((\sqrt{\lambda
   }+\epsilon ) (-8 (\lambda +3)+\epsilon ^2+8 \sqrt{\lambda }
   \epsilon )),
\eea

\bea
C_2 &=& (2 \epsilon  (2 \sqrt{\lambda } \epsilon  (9 f_1+2
   f_2+f_3)+3 f_1 \epsilon ^2-24 f_1\nonumber \\
&&{}+8 \lambda
   (f_2+2 f_3)-4 f_2 \epsilon ^2+f_3 \epsilon
   ^2+24 f_3))/\nonumber \\
&&((\sqrt{\lambda }+\epsilon ) (-8
   (\lambda +3)+\epsilon ^2+8 \sqrt{\lambda } \epsilon )),
\eea
\bea
C_3 &=& \frac{4 \epsilon  \left(\epsilon  (-f_1+2 f_2-f_3)-12
   \overline{f} \sqrt{\lambda }\right)}{-8 (\lambda +3)+\epsilon ^2+8
   \sqrt{\lambda } \epsilon },
\eea

\bea
C_4 &=& (2 \sqrt{\lambda } \epsilon  (\epsilon ^4 (75 f_1^2+2
   f_1 (58 f_2-5 f_3)+48 f_2^2\nonumber \\
&&{}+116 f_2
   f_3+75 f_3^2)-480 \epsilon ^2 (5 f_1^2+5
   f_1 f_2\nonumber \\
&&{}+2 f_1 f_3+2 f_2^2+5 f_2
   f_3+5 f_3^2)+34560 \overline{f}^2)\nonumber \\
&&{}+3 \lambda
   (\epsilon ^4(273 f_1^2+232 f_1 f_2+226
   f_1 f_3+48 f_2^2\nonumber \\
&&{}+232 f_2 f_3+273
   f_3^2)-160 \epsilon ^2 (7 f_1^2+15 f_1
   (f_2\nonumber \\
&&{}+2 f_3)+4 f_2^2+15 f_2 f_3+7
   f_3^2)+11520 \overline{f}^2)\nonumber \\
&&{}+8 \lambda ^{3/2} \epsilon
    (\epsilon ^2 (40 f_1^2+91 f_1 f_2+400
   f_1 f_3+16f_2^2\nonumber \\
&&{}+91 f_2 f_3+40
   f_3^2)-240 \overline{f} (4 f_1+f_2+4
   f_3))\nonumber \\
&&{}+4 \epsilon ^6 (2 f_1^2+2 f_1
   f_2-f_1 f_3+8 f_2^2+2 f_2 f_3+2
   f_3^2)\nonumber \\
&&{}-80 \epsilon ^4 (5 f_1^2+10 f_1
   f_2-2 f_1 f_3+8 f_2^2+10 f_2f_3\nonumber \\
&&{}+5
   f_3^2)+16 \lambda ^2 \epsilon ^2 (48 f_1^2+16
  f_1 (f_2+f_3)+3 f_2^2\nonumber \\
&&{}+16 f_2 f_3+48
   f_3^2)+960 \epsilon ^2 (7 f_1^2+2 f_1 (4
  f_2+f_3)\nonumber \\
&&{}+4 f_2^2+8 f_2 f_3+7
   f_3^2))/\nonumber \\
&&(60 (\sqrt{\lambda }+\epsilon )^2
   (-8 (\lambda +3)+\epsilon ^2+8 \sqrt{\lambda } \epsilon )^2),
\eea
and where $\overline{f}=\frac{1}{3}(f_1+f_2+f_3)$. 
%--------------------------------------------%

\bibliography{Master,refers}

\end{document}